\documentclass[12pt,preprint]{aastex}

\newcommand{\kms}{km~s$^{-1}$}
\newcommand{\lsim}{\hbox{ \rlap{\raise 0.425ex\hbox{$<$}}\lower 0.65ex\hbox{$\sim$} }}
\newcommand{\gsim}{\hbox{ \rlap{\raise 0.425ex\hbox{$>$}}\lower 0.65ex\hbox{$\sim$} }}

\shorttitle{Cl1604 Structure}
\shortauthors{Gal et al.}

\begin{document}
\title{The Complex Structure of the Cl 1604 Supercluster at $z\sim0.9$}

\author{R.~R. Gal\altaffilmark{1}, B.~C. Lemaux\altaffilmark{2}, L.~M. Lubin\altaffilmark{2}, D. Kocevski\altaffilmark{2} \& G.~K. Squires\altaffilmark{3}}

\affil{University of Hawai'i, Institute for Astronomy, 2680 Woodlawn Dr., Honolulu, HI 96822}
\altaffiltext{1}{rgal@ifa.hawaii.edu}
\altaffiltext{2}{Department of Physics, University of California -- Davis, One Shields Avenue, Davis, CA 95616} 
\altaffiltext{3}{California Institute of Technology, M/S 220-6, 1200 E. California Blvd., Pasadena, CA 91125}

\begin{abstract}

The Cl1604 supercluster at $z=0.9$ is one of a small handful of such
structures discovered in the high redshift universe, and is the first
target observed as part of the Observations of Redshift Evolution in
Large Scale Environments (ORELSE) Survey. To date, Cl1604 is the
largest structure mapped at $z\sim1$, with the most constituent
clusters and the largest number of spectroscopically confirmed member
galaxies. In this paper we present the results of a spectroscopic
campaign to create a three-dimensional map of Cl1604 and to understand
the contamination by fore- and background large scale structures.
Combining new Deep Imaging Multi-object Spectrograph observations with
previous data yields high-quality redshifts for 1,138 extragalactic objects in a
$\sim 0.08~ $deg$^2$ region, 413 of which are supercluster members.
We examine the complex three dimensional structure of Cl1604,
providing velocity dispersions for eight of the member clusters and
groups. Our extensive spectroscopic dataset is used to examine
potential biases in cluster velocity dispersion measurements in the
presence of overlapping structures and filaments. We also discuss
other structures found along the line-of-sight, including a filament
at $z=0.6$ and two serendipitously discovered groups at
$z\sim1.2$.
\end{abstract}


\section{Introduction}

Superclusters are complex structures, consisting of multiple galaxy
clusters and groups connected by chains of galaxies. They can reach
sizes of $\sim100~h^{-1}_{70}$ Mpc, making them the largest structures
in the universe. As such, they can be used to provide an estimate of
the baryonic mass fraction $\Omega_b$. Their frequency and topology
may be used to test the veracity of large scale cosmological
simulations. Because they contain structures spanning a wide range of
projected and local densities, superclusters are ideal sites for
studying the variety of physical processes affecting galaxy evolution,
including ram pressure stripping, mergers, tidal encounters,
harassment, etc. They are also sites of cluster-cluster
interactions, allowing us to probe how cluster-scale interactions
effect their constituent galaxies, and to probe possible differences
between the dark and baryonic matter distributions.

Recent deep, wide-field imaging campaigns have begun to reveal an
increasing number of superclusters at $z>0.7$. These include a $z=0.9$
compact supercluster found in the Red-sequence Cluster Survey
\citep{gil08}, a structure at $z=0.89$ in the UK Infrared Deep Sky
Survey (UKIDSS) Deep eXtragalactic Survey \citep{swi07}, and a large
galaxy density enhancement at $z\sim0.74$ in the COSMOS field
\citep{sco07}. To increase the number of well-studied high-redshift
structures, we have begun the Observations of Redshift Evolution in
Large Scale Environments (ORELSE) survey, a systematic photometric and
spectroscopic search for structure on scales $>10$ Mpc around 20 known
clusters at $z> 0.6$ \citep[][hereafter ORELSE I]{lub08} . The Cl 1604
supercluster at $z\sim0.9$  was the first supercluster detected as part
of this survey and remains the most heavily studied structure at these
redshifts. It was initially discovered as two separate clusters in the
plate-based survey of \citet{gho86}, with redshifts and preliminary
velocity dispersions measured by \citet[][hereafter P98 and
P01]{pos98,pos01}. Deeper CCD imaging \citep{lub00} revealed a total
of four distinct galaxy density peaks in a contiguous area of $10.4'
\times 18.2'$. A large spectrographic survey using the Low-Resolution
Imaging Spectrograph \citep[LRIS;][]{oke95} and the Deep Imaging
Multi-object Spectrograph \citep[DEIMOS;][]{fab03} on the Keck 10-m
telescopes confirmed the new cluster candidates and provided improved
velocity dispersions for all four supercluster components using 230
galaxies \citep{gal04}. The large radial depth of the supercluster
($\sim93~h^{-1}_{70}$ Mpc) in comparison to the small imaging area
($4.8\times8.5~h^{-1}_{70}$ Mpc) spurred a wide field imaging campaign
using the Large Format Camera \citep[LFC;][]{sim00} on the Palomar 5-m
telescope. Four additional red ($1.0<=r'-i'<=1.4$) galaxy
overdensities, identified as candidate clusters, were detected in the
larger area, bringing to eight the total number of possible clusters
in this structure \citep{gal05}.

Additional multi-object spectroscopy was undertaken to confirm the new
cluster candidates and improve our understanding of the supercluster
structure. In this paper we utilize our complete spectroscopic data
set, with $\sim1100$ extragalactic redshifts, to examine the three
dimensional structure of the Cl1604 supercluster, improve our cluster
detection algorithm, and study the cluster galaxy populations. A total
of 427 galaxies have been spectroscopically confirmed within the
structure, nearly doubling the number of known members. The only
comparable structures at similar redshift are the $z=0.89$ UKIDSS
supercluster with five density peaks \citep{swi07} and the $z=0.9$
three-cluster system described in \citet{gil08}.  These two systems
each have only $\sim50$ spectroscopic members. We show that even with
hundreds of members spread over the supercluster, it remains difficult
to obtain consistent velocity dispersions due to the presence of
overlapping structures, filaments, and changes in galaxy populations
between clusters.

Section 2 describes the photometric data, density mapping and cluster
detection, all of which are improvements on our earlier work. The
spectroscopic data is described in \S3. Cluster velocity dispersions
and the associated measurement complications, along with the overall
structure of the supercluster, are discussed in \S4.  We utilize our
extensive spectroscopic sample to study photometric selection of
cluster members using only three filters ($r', i', z'$), and the
cluster properties. Section 5 details the various fore- and background
structures, including two serendipitously discovered structures at
$z>1.1$ and an apparent wall at $z\sim0.6$. A brief discussion of implications
for optical surveys of high-redshift galaxy clusters and
spectroscopic studies of cluster galaxy populations, is presented in
\S6. Throughout the paper we use a cosmology with $H_0=70$ km s$^{-1}$
Mpc$^{-1}$, $\Omega_m=0.3$ and $\Omega_{\Lambda}=0.7$.

\section {Imaging and Cluster Detection}
The imaging data are presented in \citet{lub00} and \citet{gal05}. We
provide a brief review, along with a description of improvements to
the photometric calibration and modifications made to the final
density map since those publications. We have also implemented a
technique to estimate the contamination of our cluster catalog by
false detections, described in \S2.4.

\subsection{Photometric Data}
The photometric data are a combination of two pointings taken in Cousins $R$
and Gunn $i$ with the Carnegie Observatories Spectroscopic Multislit
and Imaging Camera \citep[COSMIC;][]{kel98} and two pointings in Sloan
$r'$,$i'$ and $z'$ with the LFC,
both on the Palomar 5-m telescope. The layout of the imaging fields is
shown in Figure 1 of \citet{gal05}, who also describe the LFC
observations and data reduction. The details of the COSMIC
observations and data reduction are described in \citet{lub00}.

Since the publication of \citet{gal05}, the Sloan Digital Sky Survey
\citep[SDSS;][]{yor00} has released data in the fields covered by
these pointings. We have used this data to improve the transformation
of the COSMIC $R$ and $i$ magnitudes to the SDSS system and to
calibrate the LFC fields. The SDSS DR5 \citep{ade07} was queried for
the $r'$, $i'$ and $z'$ model magnitudes in the region covered by our
imaging. Objects were cross-identified between the SDSS catalog and
our four separate pointings (two COSMIC and two LFC). The scatter
between our coordinates and those of SDSS were $\sim0.25''$ in both
right ascension and declination, commensurate with the expected
astrometric errors. Transformations from our photometry to SDSS were
derived independently for each filter and pointing. The COSMIC $R$ and
$i$ magnitudes were converted to the SDSS system using equations of
the form
\begin{equation}
r'_{SDSS} = A\times R + B\times (R-i) + C
\end{equation}
For objects detected in only one band, we assumed $R-i=0.71$, the median color 
of sources in the COSMIC catalog. Similarly, the LFC data are recalibrated using
\begin{equation}
r'_{SDSS} = A\times r'_{LFC} + B\times (r'_{LFC}-i'_{LFC}) + C
\end{equation}

The depths of the LFC images are such that any object with normal
colors will be detected in all three bands, unless it as the field
edges or near a saturated star. We therefore apply no color terms to
single-band detections. There is significant overlap between the two
LFC pointings, and especially the COSMIC and LFC pointings. In our
previous work, we simply trimmed the catalogs to non-overlapping areas
based on visually selected coordinate limits. Here, we have improved
the catalog combination by selecting objects based on photometric
errors. First, the two LFC pointings were compared to each other. For
sources appearing in both pointings, we use the detection with the
lowest average photometric error in the three filters to create a
master LFC catalog. The same procedure is applied to the two COSMIC
catalogs to generate a single COSMIC catalog. These two catalogs are
then compared on the basis of the $r'$ and $i'$ errors, with objects
selected from LFC if the average photometric error
$(err_{r'}+err_{i'})/2$ from LFC is less than that from COSMIC plus
0.05 mag. This scheme gives preference to data from the LFC
because it covers a larger area and includes photometry in $z'$. For
sources where COSMIC $r'$ and $i'$ photometry is chosen, we
incorporate $z'$ data if it is available from LFC, or from SDSS if
there is no LFC $z'$ data or the LFC $z'$ detection is flagged as bad. The
final hybrid catalog is corrected for Galactic reddening using the
dust maps from \citet{sfd98} on an object-by-object basis. The
$5\sigma$ limiting magnitudes are 25.2, 24.8, and 23.3 in $r'$, $i'$
and $z'$, respectively.

\subsection{Producing the Density Map}
\begin{figure}
\plottwo{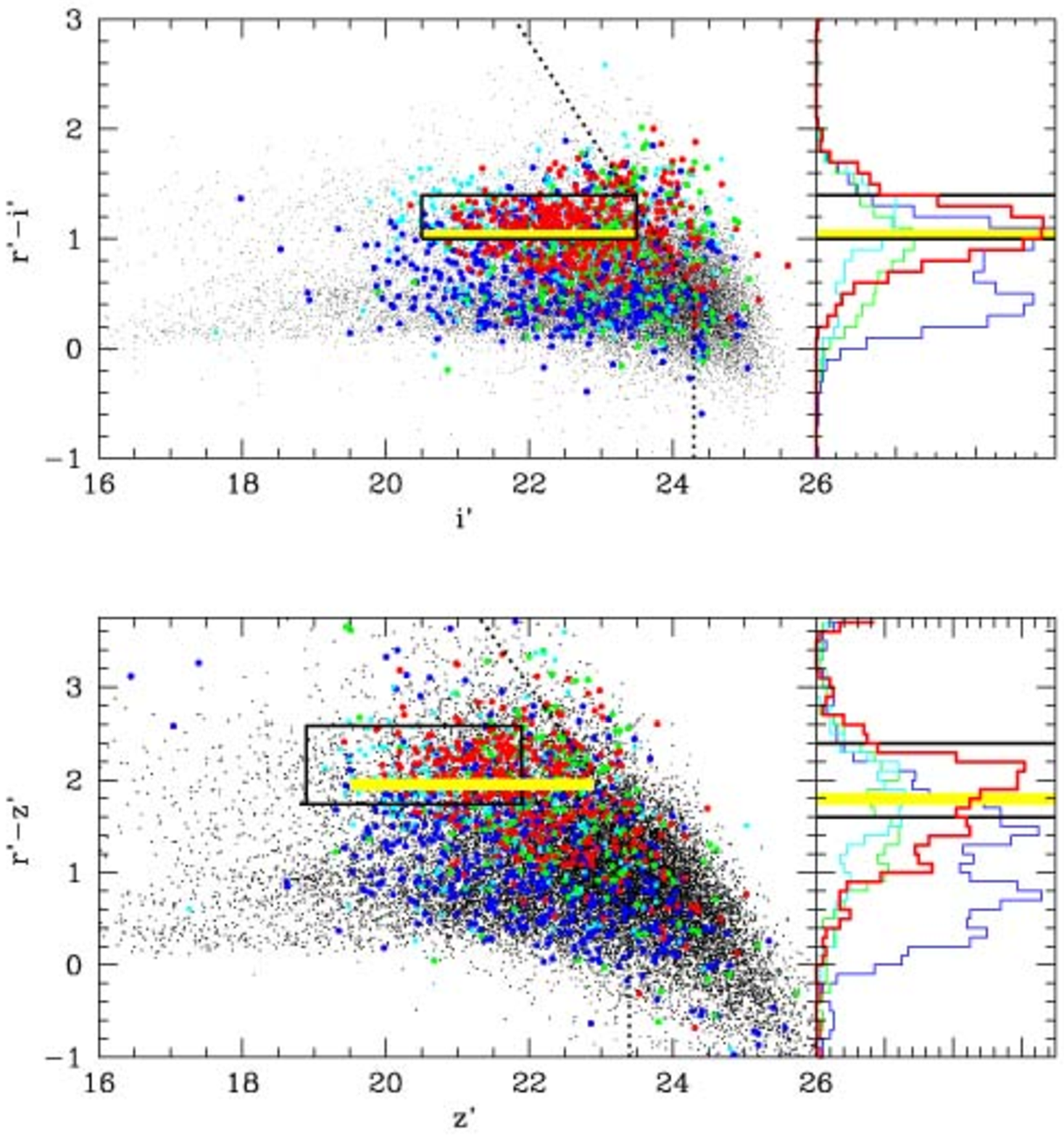}{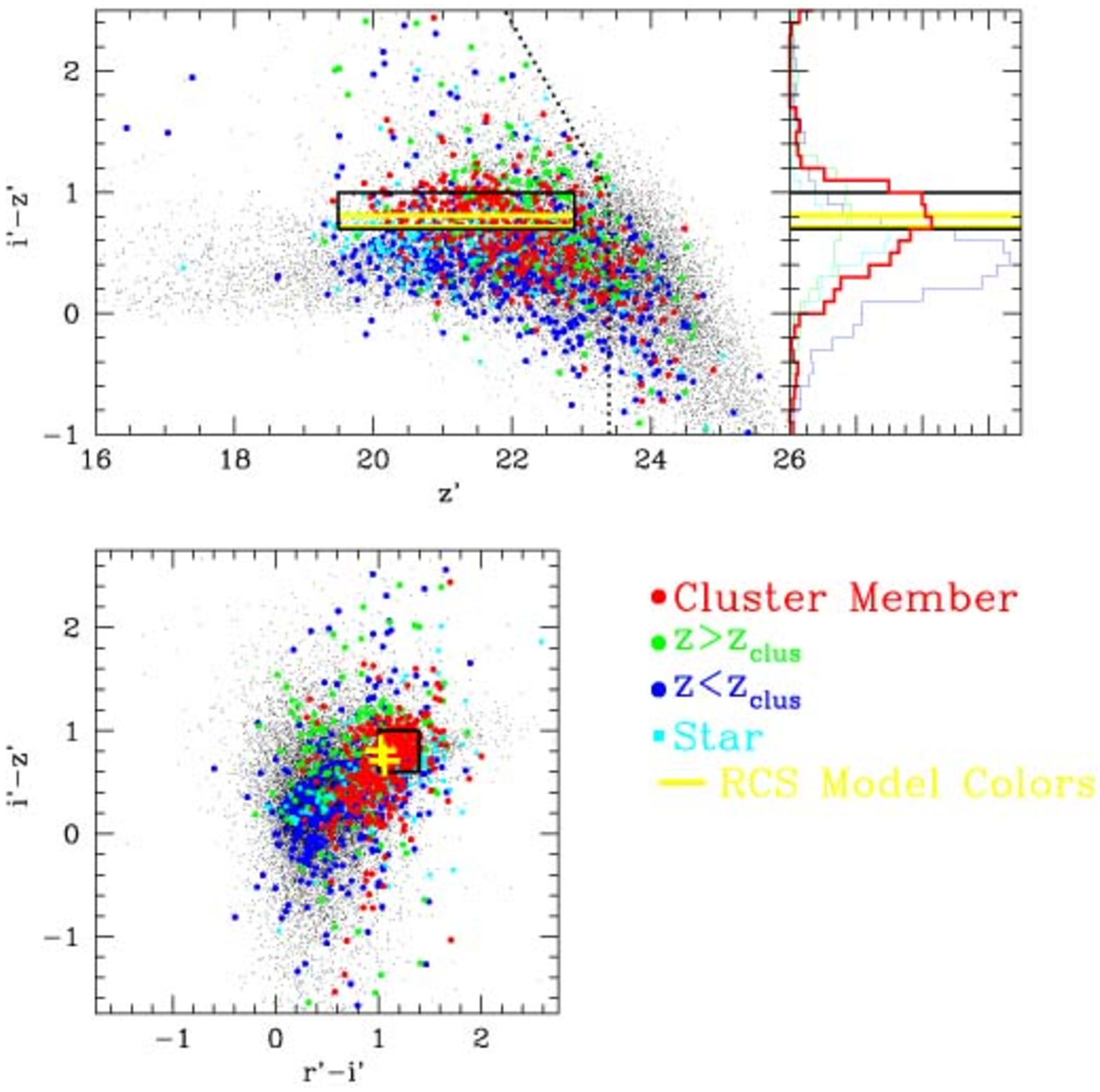}
\caption{The $i'$ vs. $(r'-i')$, $z'$ vs. $(r'-z')$, $z'$ vs. $(i'-z')$ and $(r'-i')$ vs. $(i'-z')$ color-magnitude and color-color diagrams of all objects in our LFC and COSMIC imaging areas. The black rectangular regions outline the color-selections applied to produce the density maps and prioritize spectroscopic targets. The yellow bars and crosses show the synthetic colors for a $z=0.9$ early-type galaxy based on the RCS models; they are within our broad color cut but not an exact fit to the observed colors. Spectroscopically confirmed supercluster members are overplotted as large red dots. Foreground galaxies at $z<0.84$ are shown as blue dots, background galaxies at $z>0.96$ are green dots, and cyan dots are stars. Color distributions of spectroscopic objects are shown to the right of each CMD.}
\label{cmds} 
\end{figure}

Following \citet{gal04}, we produce a galaxy density map by adaptively
smoothing a color-selected subset of galaxies in the Cl1604
field. With the additional spectroscopy presented in \citet{gal05} and
the improved photometric calibration discussed above, some
modifications to the density mapping algorithm were made. First, we
used the observed colors of $\sim300$ confirmed members to choose
intervals in both the $r'-i'$ and $i'-z'$ colors that provided large
numbers of supercluster galaxies and enhanced contrast relative to
fore- and background structures. The limiting colors and magnitudes
used were $1.0\le(r'-i')\le1.4$, $0.6\le(i'-z')\le1.0$ and
$20.5\le i' \le23.5$. These color cuts are delineated by the black rectangles in the
color-magnitude and color-color diagrams in Figure~\ref{cmds}. 

We also examined the possibility of applying color cuts based on
stellar population synthesis models. Synthetic galaxy spectra were
generated following the prescription used by the Red-sequence Cluster
Survey \citep[RCS,][]{gla05}, with a \citet{bru03} model parameterized
as a 0.1 Gyr burst ending at $z=2.5$ followed by a $\tau=0.1$ Gyr
exponential decline. This model predicts colors of $(r'-i')=1.05$ and
$(i'-z')=0.78$ at $z=0.9$; they are plotted as the yellow bars and
crosses in Figure~\ref{cmds}. The observed $(r'-i')$ colors are about
0.2mag redder. Although the BC03 model colors are within our broad
color cuts, the empirical colors, by construction, provide optimal
detection of structure at the redshift of interest. Tests of other
models for the red galaxy population, including single starbursts with
either instantaneous or exponentially declining bursts, showed that
expected galaxy colors varied a few tenths of a magnitude between
models. Therefore, we use our empirically confirmed color cuts to
produce a final density map. As discussed in ORELSE I, the star formation history chosen
by the RCS yields model colors that, while within our broad color
cuts, do not match perfectly the observed red sequences in any of our
structures at $0.7<z<1.1$. This is unsurprising as the average
galaxy's star formation history varies with environment, and there
remain unresolved issues with stellar population synthesis models. As
models improve and large surveys like ORELSE, RCS and PISCES
\citep[Panoramic Imaging and Spectroscopy of Cluster Evolution with
Subaru,][]{kod05} obtain spectroscopy and map out the observed red
sequence colors as a function of redshift, it will be possible to find
BC models that better fit the data.

\begin{figure}
\plotone{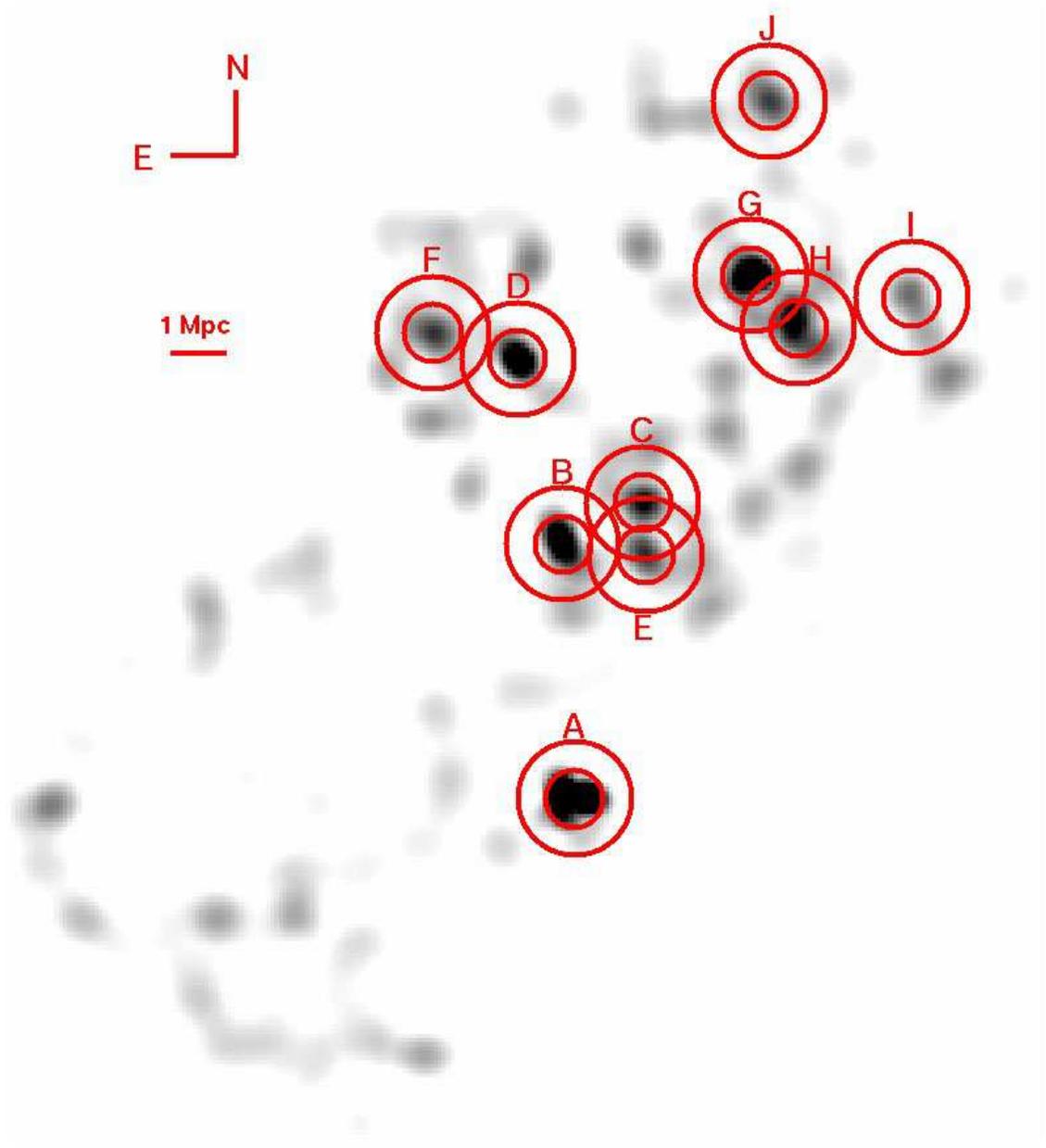}
\caption{Adaptive kernel density map of color-selected galaxies in the Cl1604 region. Detected cluster and group candidates are marked with red circles of $0.5h^{-1}_{70}$ and $1.0h^{-1}_{70}$ radius, and labeled with an identifying letter.}
\label{akmap} 
\end{figure}

The color and magnitude cuts select only 722 objects out of an
initial catalog of $\sim12,000$ with $i'<23.5$. An adaptive kernel
smoothing is applied using an initial window of $0.75 h^{-1}_{70}$ Mpc
and 10 arcsecond pixels. We use a smaller kernel than the
$1h^{-1}_{70}$ Mpc radius applied in \citet{gal04} because our extensive
spectroscopy showed that small groups were being blended into single
detections in the density map. The smaller kernel also enhances the
contrast of small groups against the background, making detection of
such low-mass systems easier. The new density maps are shown in
Fig.~\ref{akmap}, with the detected overdensities marked. Comparison
to Figure 1 of \citet{gal04} shows that similar structures are found
with the new color cuts.

\subsection{Cluster Detection}

The density maps are used primarily to provide a visual locator for
intermediate density large scale structure in the imaging
field. Filaments and cluster infall regions can cover significant
portions of the observed area, but at relatively low contrast, making
it difficult to detect them and define their boundaries. Identification
of such overdense regions is used primarily to direct placement of follow-up
observations, especially slitmasks for multi-object
spectroscopy. However, it is necessary to provide a consistent
detection mechanism for finding clusters and groups in the ORELSE
survey fields, described below.

Cluster detection is performed by running SExtractor \citep{ber96} on
the adaptively smoothed red galaxy density map. In \citet{gal04} we
relied on visual inspection of the density map and comparison to the
poorest (lowest velocity dispersion) cluster in the field, Cl1604+4316
(cluster C). With the addition of the spectroscopy presented in
\citet{gal05} and herein, the projected overdensities can be better
characterized, and detection parameters tuned to detect real
structures. We tested various combinations of detection threshold and
minimum area, all taken from Cluster C. This cluster still has the lowest reliably measured velocity dispersion, as detailed in \S4. SExtractor was run with these
parameters, and the resulting detections compared to our spectroscopic
cluster catalog to ensure that all confirmed clusters and groups were
detected. We also examined the false detection rate, described in the
following section. The final set of detection parameters has high
detection efficiency (100\% by design) for the confirmed structures
while yielding less than one false detection on average in the survey
area. 

A total of 10 cluster and group candidates are detected, similar to
those found in \citet{gal04}. Where possible, we use the same naming
and lettering convention as our previous work to refer to specific
clusters. Table~\ref{candinfo} provides the details of each
cluster. Column 1 lists the single letter denoting each cluster \citep[following the nomenclature of][]{gal05}, while
Column 2 provides the full name. Columns 3 and 4 give the updated
cluster coordinates. The remaining columns provide information on
membership, redshifts, and velocity dispersions, as detailed in \S4.

\subsection{Cluster Detection Thresholds}

Because one of our goals is the detection of low-mass group-like structures in the
Cl1604 field, our detection thresholds are generous in the sense that
we may suffer from a high false detection rate, even with the
stringent color cuts. Our method for setting thresholds and estimating
contamination rates is detailed in ORELSE I; however we outline
it briefly here since the Cl1604 data have been used to tune the
parameters.

\begin{figure}
\plotone{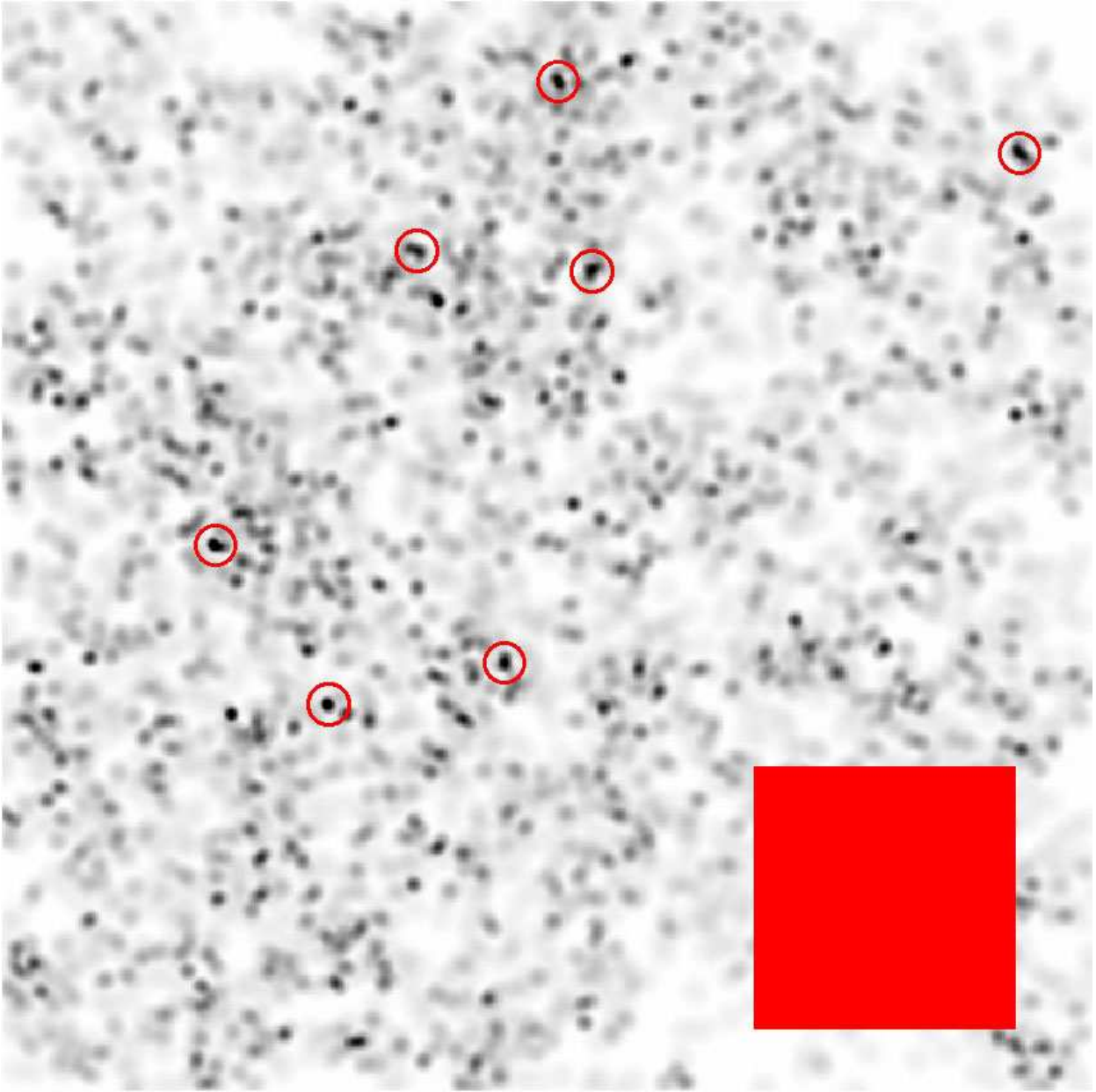}
\caption{Adaptive kernel density map of a Raleigh-Levy simulated galaxy distribution covering $2.2^{\circ}$ on a side. Cluster candidates, which would be false detections since there are no real structures in this map, are marked with circles of $1.0h^{-1}_{70}$ radius (for $z=0.9$). The shaded box in the bottom right covers an area equal to that imaged for Cl1604.}
\label{rlfig} 
\end{figure}

We use the NOAO Deep Wide Field Survey \citep{jan99} to study possible
contamination by 'field' galaxies having colors meeting our selection
criteria. We cannot rely on our imaging survey because it is entirely
targeted at known high-density regions of the universe. Thus, we use
the NOAO DWFS to derive the statistical properties of the red galaxy
distribution. The third data release (DR3) in the Bootes field covers
an area of nine square degrees, in the B$_W$, R and I filters. In
brief, we transform the DWFS $R$ and $I$ data to the SDSS $r'$ and
$i'$ systems, and apply our $r'-i'$ color cut and $i'$ magnitude
limit. Because there is no $z'$ photometry in the DWFS, we then assume
that field galaxies meeting only the $r'-i'$ cut (numbering
$N_{1color,DWFS}$) have a similar projected distribution as those that
also meet our $i'-z'$ cuts, only with a higher space density. We
tested this assumption by examining the galaxy distributions far from
cluster centers in density maps using our own data, applying only the
$r'-i'$ cut and cuts in both colors. We find no significant
differences. Following \citet{pos96} and \citet{gal04}, we use the
color- and magnitude-selected subset of the DWFS data to compute
Raleigh-Levy (RL) parameters for galaxies meeting our color cuts,
which we then use to generate simulated galaxy distributions. These
are then used to produce density maps on which we run SExtractor. The
SExtractor parameters DETECT\_THRESH and MIN\_AREA ($\pi\times
r_{test}^2$) are varied, using values of the galaxy density at
different radii $r_{test}$ from the center of cluster C in Cl1604,
until a set of values is found that successfully detects the confirmed
clusters in Cl1604 while yielding low contamination. We find that an
area of 47 pix$^2$, corresponding to a radius of 0.3$h^{-1}$ Mpc,
allows us to detect all of the spectroscopically confirmed cluster
candidates in the real data, while minimizing the number of detections
in the RL simulations, which should all be chance projections. The
median number of false detections expected in the area imaged for
Cl1604 is $N_{false}=1.4$, compared to the ten candidates detected.
Figure~\ref{rlfig} shows an example RL simulated map $2.2^{\circ}$ on
a side, with the eight 'false' candidates detected using our optimal
SExtractor parameters marked. The shaded box in the bottom right
covers an area equal to that imaged for Cl1604.

\section{Spectroscopic Data}
\renewcommand{\labelitemi}{}
\renewcommand{\labelitemii}{}

The spectroscopic observations consist of five separate datasets:
\begin{enumerate}
\item Spectra taken using LRIS as part of the original \citet{oke98} survey.
\item Additional LRIS spectra taken in May 2000, covering the area imaged with COSMIC.
\item DEIMOS spectra covering the same area taken in 2003.
\item DEIMOS spectra taken in 2005 and 2006, using color selection and spanning a larger area imaged with LFC.
\item DEIMOS spectra covering the same area as before, but including X-ray, radio, HST ACS and new color-selected sources.
\end{enumerate}
Each sample is described in more detail below, with the layout of all masks shown in Fig.~\ref{masks}.

\subsection{LRIS and DEIMOS Observations}

The original observations of Cl1604+4304 and Cl1604+4321 (clusters A
and D) were taken in the late 1990s with LRIS on the Keck I and II
telescopes and are detailed in \citet{oke98}, \citet{pos98} and
\citet{lub98}. These initial slitmasks targeted {\em all} galaxies
with $R<23.5$ based on earlier imaging with LRIS. It is important to
note that no color selection was utilized in selecting galaxies to
observe; therefore, blue galaxies at the supercluster redshift are
more likely to be observed compared to the DEIMOS data discussed
below. Redshifts were measured for 103 and 135 galaxies in the fields
of Cl1604+4304 and Cl 1604+4321, respectively.  Imaging of a larger
region with COSMIC on the Palomar 200-inch telescope was used to
design additional LRIS slitmasks, targeting an area including clusters
B and C. Objects with magnitudes down to $i=23.0$ were included, with
priority given to red galaxies. Six additional LRIS masks with 156
targets were observed in May 2000. Median velocity errors for the LRIS
data are 150~{\kms}.

Starting in May 2003, Cl1604 was observed with DEIMOS on the Keck II
telescope. We briefly describe these observations here; the details
will be provided in a future paper releasing multi-wavelength data,
including redshifts, in the Cl1604 field.  The first
DEIMOS observations also spanned the region imaged with COSMIC, and
are detailed in \citet{gal04}. Galaxies as faint as $i\sim24$ were
observed, with higher priority for red galaxies based on COSMIC $R-i$;
only 60\% of the targeted objects met the red galaxy color cut. We
used the 1200 l/mm grating, blazed at 7500\AA, and $1''$ slits,
resulting in a pixel scale of 0.33\AA~pix$^{-1}$, a resolution of
$\sim$1.7\AA\ (68 km s$^{-1}$), and typical spectral coverage from
6385\AA~to 9015\AA. Data were reduced using the DEEP2 version of the
{\em spec2d} and {\em spec1d} data reduction pipelines
\citep{dav03}. Since the publication of \citet{gal04}, the DEEP2 team
has made available their redshift measurement pipeline {\em zspec}
\citep{coo07} and all DEIMOS redshifts have now been obtained using this
package. Typical redshift errors using this method are 25~{\kms},
nearly a factor of four improvement over our original measures.

\begin{figure}
\plotone{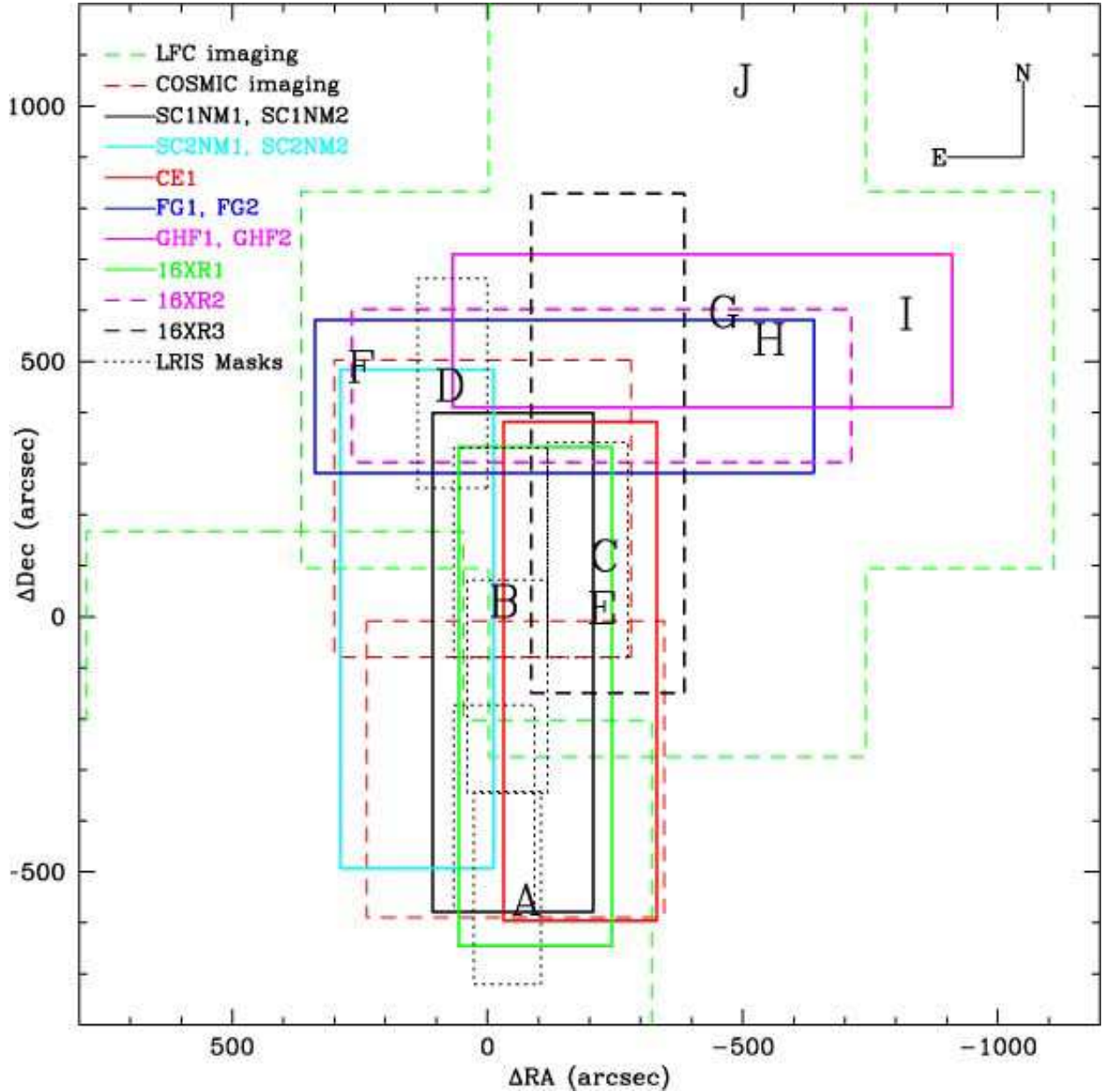}
\caption{Layout of spectroscopic observations in the Cl1604 field. The regions imaged from the ground with COSMIC (red dashed lines) and LFC (green dashed lines) are shown, along with the LRIS masks observed in May 2000 as dotted lines. The eleven DEIMOS masks from the three campaigns discussed in the text 
are also plotted, along with the locations of the clusters detected from the density map.}
\label{masks} 
\end{figure}

Using the multiband LFC photometry from \citet{gal05}, galaxies with
$20.5\le i' \le 24.0$ were targeted over a larger area based on both
color and magnitude.  Five DEIMOS masks, labeled CE1, FG1, FG2 and
GHF1 and GHF2 in Fig.~\ref{masks} were observed in 2005 and 2006 using
these ground-based color selection criteria.  The Cl 1604 supercluster
was also the subject of a two-band 17-pointing mosaic with the
Advanced Camera For Surveys (ACS), a two-pointing ACIS-I observation
with Chandra \citep{koc08}, and a Very Large Array B-Array 1.4 Ghz
radio map \citep{mil08}. These multi-band data were used to design
additional DEIMOS masks. The primary sample included X-ray and radio
sources within the ACS mosaic, and red-sequence galaxies selected on
the basis of the high-precision ACS colors. Secondary samples included
X-ray and radio sources outside the ACS mosaic and bluer galaxies
within the ACS mosaic. Lower priority samples include red galaxies
outside the ACS mosaic, then bluer (by as much as 0.3$^{\rm mag}$) galaxies
outside the ACS area, and finally any galaxy not already selected by
the above criteria. We observed three of these masks in June 2007.

Figure~\ref{masks} shows the positions of the LRIS masks observed in
May 2000 as well as all eleven DEIMOS masks. The locations of the
cluster and group candidates are also marked. This figure makes
evident the uneven spectroscopic coverage of the
supercluster. Especially noteworthy is the extensive LRIS coverage of
Clusters A and D; because no color selection was used for the original
LRIS masks, the blue (presumably star-forming) populations in these
clusters will be better sampled. Nevertheless, even in regions where
there is only DEIMOS coverage, $\sim50\%$ of the targeted galaxies
were outside the red sequence. We explore the dependence of cluster
velocity dispersions on the population sampling in \S4.1.3.

In addition to the targeted objects, numerous serendipitous spectra
were found by visual inspection. These objects had their extraction
windows set manually, but were otherwise run through the same
pipelines as the target objects. Two of us (RG and BL) systematically
inspected all of the {\em spec2d} created serendip one-dimensional
spectra to determine whether these extractions contained genuine
serendips and to confirm their redshifts.  A significant number of
serendips (180 extragalactic objects, 138 with Q$>2$) were found in
our DEIMOS data. Many of these at the supercluster redshift show
strong [OII] emission and little continuum. These objects represent a
different population than the primary targets (red sequence galaxies)
and may impact the velocity dispersion determination, explored in
\S4.1.3.

\subsection{Combined Spectroscopy Results}

All of the spectroscopic data were combined into a single master
catalog. For objects observed with both LRIS and DEIMOS, we used the
DEIMOS results due to the improved redshift accuracy with the higher
dispersion grism and typically higher signal-to-noise
ratio. Serendipitous spectra were matched to photometric objects from
ACS and/or Palomar imaging. In a few cases no counterpart was found,
usually for single emission line spectra, some of which are
likely to be Ly$\alpha$ emitters at $z>4$ \citep{lem08a}. Each photometric
identification was examined visually (by RG) to check for blends,
interacting galaxies and misidentifications. A set of flags was added
to the catalog, identifying whether or not the object was detected
and/or blended in either or both the ACS and Palomar images. The final
catalog contains a total of 1,671 unique objects. Of these,

\begin{list}{}{\setlength{\itemindent}{-20pt} \setlength{\itemsep}{0pt}}
\item 1,215 are extragalactic objects, of which 1,138 have Q=3 or Q=4;
\item 1,148 (1,089 with Q=3 or 4) of the above are matched to a cleanly detected (not blended) photometric object in either the ACS or ground-based imaging;
\item 140 are stars;
\item 427 are in the redshift range of the supercluster ($0.84 \le z \le 0.96$), of which 413 have Q=3 or Q=4;
\item 417 (404 with Q=3 or 4) of the supercluster members have clean photometry.
\end{list}

\begin{figure}
\plotone{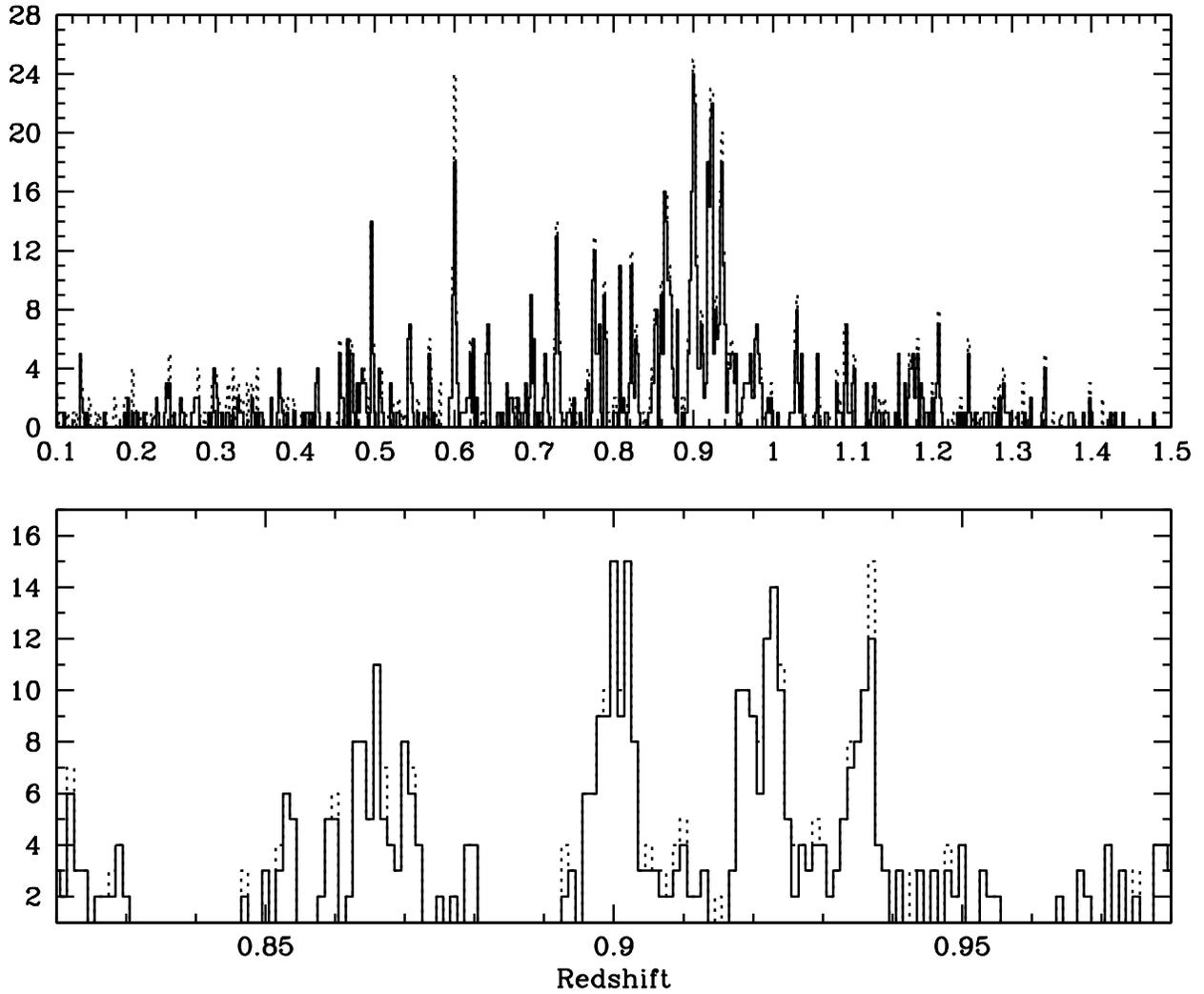}
\caption{Redshift distribution in the Cl1604 field. The top panel shows all extragalactic objects (dotted line) and only those with high redshift quality (solid line). The bottom panel focuses on the supercluster redshift range with redshift bins of $\Delta z=0.001$.}
\label{spechist} 
\end{figure}

The top panel of Figure~\ref{spechist} shows the redshift distribution
of the Q=3 or 4 extragalactic objects (solid line), as well as for
objects with Q$>2$ (dotted line). The bottom panel
shows only the redshift range of the supercluster ($0.84\le z \le
0.96$) with redshift bins of $\Delta z=0.001$. In addition to a number
of clear redshift peaks in the supercluster, many intervening
structures are noticeable, discussed in detail in \S5. We use only
the most reliable (Q=3 or 4) redshifts for all of the analyses
presented here.

\section{Cluster Properties and Supercluster Structure}

\subsection{Individual Cluster Properties}

Nine of the ten cluster candidates listed in Table~\ref{candinfo} have
been extensively mapped in our spectroscopic campaign, with only
Cluster J lacking spectroscopic coverage. We use our dataset to
examine the redshift distribution around each cluster candidate, both
to verify the structures and measure velocity dispersions.  Because of
our generous detection threshold, we expect to find structures
consistent with modest sized groups as well as richer clusters. In
some cases, clusters overlap each other even with the smallest radius
used, and almost all clusters have companions within 1 $h^{-1}_{70}$
Mpc, as seen in Fig.~\ref{akmap}. The complex supercluster structure
makes separation of individual cluster components difficult and in
some cases impossible. Information on both position and velocity must
be incorporated; even so, individual clusters or groups may exhibit
significant substructure \citep{dre88}.

\subsubsection{Velocity Dispersions}
\begin{figure}
\plotone{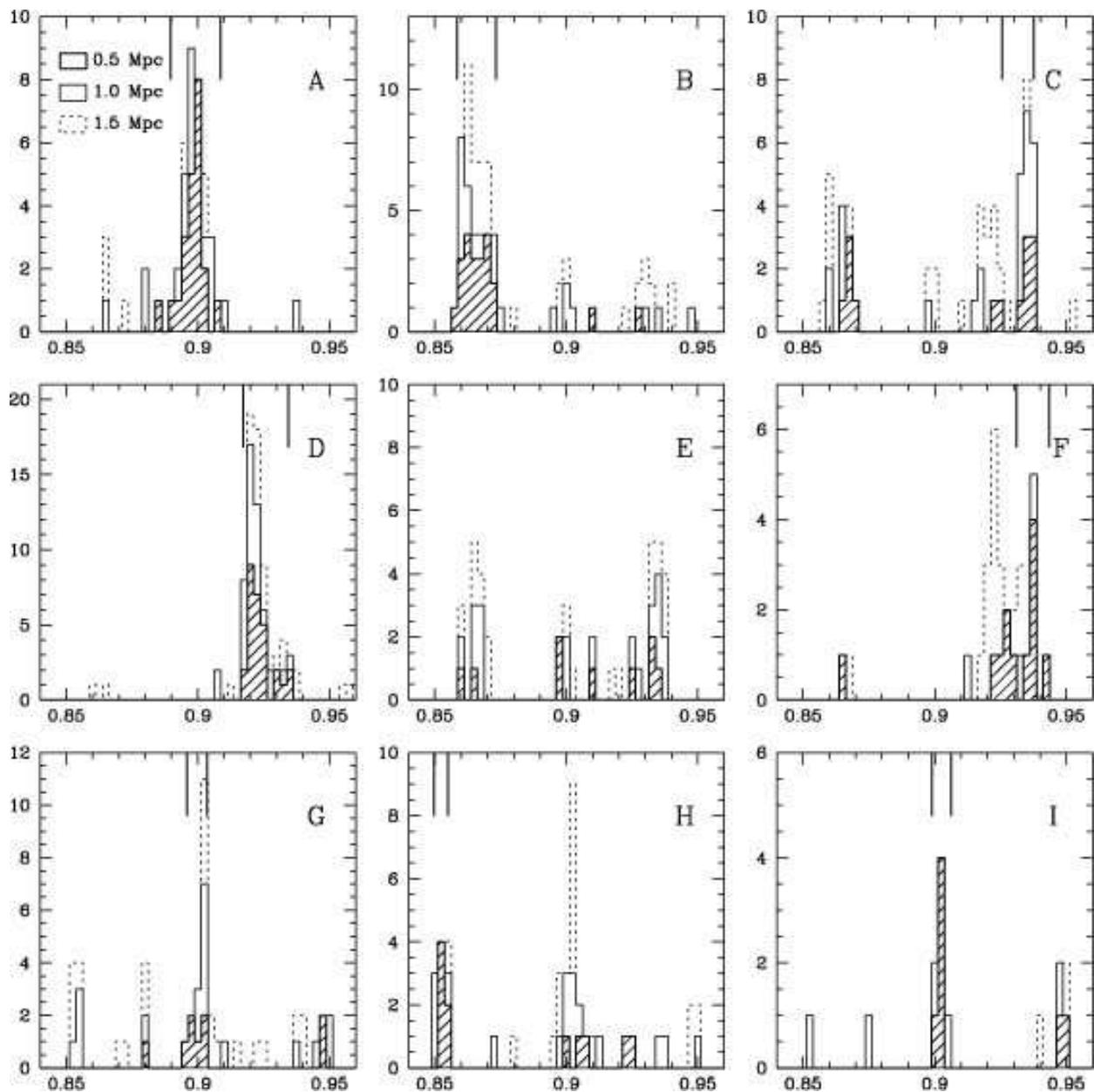}
\caption{Redshift histograms for Clusters A-I. The shaded regions show 
the distribution within a projected radius of $0.5 h^{-1}_{70}$ Mpc,
while the solid and dashed lines correspond to radii of 1 and 1.5
$h^{-1}_{70}$ Mpc, respectively. The vertical lines at the top of each
panel demarcate the final redshift range used to compute the velocity
dispersion for each cluster. Note that the vertical axes have
different ranges. Cluster E is not marked due to contamination from B and C.}
\label{velhists} 
\end{figure}

We use the cluster centers determined by SExtractor from the density
map when measuring cluster redshifts and velocity dispersions. We note
that the spectroscopic coverage varies greatly from cluster to
cluster, as seen in Fig.~\ref{masks}.  We first construct redshift
histograms within 0.5, 1.0 and 1.5 $h^{-1}_{70}$ Mpc projected radii
centered on each cluster. These are plotted in Figure~\ref{velhists}
for clusters A-I. The shaded region, solid line and dashed line show
the redshift distributions within 0.5, 1.0 and 1.5 $h^{-1}_{70}$ Mpc,
respectively. An initial redshift range (typically spanning $\sim3000$ km
s$^{-1}$) for each cluster is then
selected by visual inspection of these plots, attempting to avoid
clear redshift peaks from adjacent structures. 

The extensive overlap
between structures results in contamination of the redshift
distributions. A good example is Cluster C ($z\sim0.935$), where we
see numerous galaxies at $z\sim0.865$ within a projected $0.5
h^{-1}_{70}$ Mpc radius. In this case, the redshift separation is
large enough to easily distinguish the two populations. More complicated
overlaps, such as between D and F, where the redshift separation is
only $\Delta z\sim 0.01$, are difficult to resolve. In such cases,
neglecting the overlap may lead to overestimation of the velocity
dispersion. Conversely, applying strict redshift limits to avoid such
overlaps can artificially lower the measured dispersion. 

For each cluster we follow the procedure described in \citet{gal05}
and \citet{lub02}. After the initial redshift range is chosen, the
cosmologically corrected velocities relative to the median cluster
redshift are calculated for each galaxy within the three radii. The
initial velocity windows are typically $\pm3000$ km s$^{-1}$,
sufficiently broad to avoid biasing the dispersion estimates to lower
values. These distributions are iteratively clipped at $3\sigma$,
where $\sigma$ is the biweight dispersion computed by ROSTAT
\citep{bee90}. Typically from 0-2 galaxies are excluded from each
cluster by this clipping. The final dispersions are computed using the
biweight estimate of the scale, with errors taken from the jackknife
confidence interval on the biweight scale. These have been shown to be
well-behaved under most circumstances for modest sample sizes
\citep{bee90}.

Contamination from fore/background clusters, groups and filaments
could bias the measured velocity dispersions and their associated
errors when using ROSTAT \citep{and07}. We have used quite stringent
spatial and redshift cuts to establish the starting ranges for
velocity dispersion measurement, which reduces the contamination. We
also examined the agreement between the $3\sigma$-clipped standard
dispersion, the biweight estimate of the scale, and the gapped
estimator, noting how many (out of 3 maximum) of these estimators
agree within the quoted errors.  The results are presented in
Table~\ref{candinfo}. Columns 5, 8 and 11 give the number of
spectroscopically confirmed members within 0.5, 1.0 and 1.5
$h^{-1}_{70}$ Mpc projected radii, respectively. These numbers include
only those galaxies used to compute the velocity dispersions within
each radius. Columns 6, 9 and 12 provide the median redshift for each
cluster within each of the three radii, while Columns 7, 10 and 13
list the computed velocity dispersions and their errors. Columns 7, 10
and 13 also include, in parentheses, the number of different
dispersion estimators that agree within the errors.  Despite having
fewer galaxies, we would prefer the final velocity dispersions to be
those measured within 0.5 $h^{-1}_{70}$ Mpc, since the smaller radius
reduces contamination from overlapping clusters. However, only
Clusters A-D have sufficient members within this small radius to
compute dispersions, so we quote the dispersions using the 1.0
$h^{-1}_{70}$ Mpc radius throughout. The final redshift intervals used
for this radius are shown in Fig.~\ref{velhists} as the vertical lines
near the top of each panel. The sole exception is Cluster E, which
appears to be a superposition of components related to Clusters B and
C. We, therefore, measure no velocity dispersion and do not show its
color-magnitude diagrams.

\subsubsection{Velocity Dispersions, Errors and the Effect of Interlopers}

If we assume that the velocity dispersion errors quoted in
Table~\ref{candinfo} are due purely to sample size, then we would
expect the errors to decrease with the square root of the number of
galaxies sampled. This is almost exactly true for Clusters A and D,
and the quoted dispersions using different radii all agree within
$1\sigma$. For Cluster B, the three quoted dispersions agree very
well, even though the estimated error remains nearly constant with
sample size. For the remaining structures, which all have much sparser
sampling and appear to be intrinsically poorer, the effects of small
numbers and contamination from interlopers make reliable dispersion
estimates impossible. A good example is Cluster G, where the estimated
error increases despite the larger sample size when moving from 1.0 to
1.5 $h^{-1}_{70}$ Mpc projected radius. We see from Fig.~\ref{masks}
that Clusters G and H are very close to each other. Clearly, even with
extensive spectroscopy it is not always possible to definitively
assign membership of individual galaxies to specific clusters or
groups. The behavior of our dispersions and associated errors shows
that systematic errors remain an important contributor, likely
due to galaxies that are not physically bound to each cluster found in
the redshift range and spatial region used in the calculation. This
suggests that dense sampling of cluster cores may be necessary to
obtain reliable velocity dispersions. However, comparison of such data
to models requires understanding of possibly cluster mass dependent
biases of the observed versus true dispersions.

Another possibility, discussed by \citet{and07}, is that the
dispersions and errors are not measured correctly due to interlopers
from filaments. The presence of an approximately uniform background
results in artificially inflated dispersions. This background
contribution is important as long as it is at least a modest 
fraction of the total sample. In \citet{and07} they show the exaggerated
case of 500 cluster members and 500 interlopers. We utilize Clusters A
and D to see if such interlopers affect significantly our dispersions.

One way to estimate the number of potential interlopers is to count
galaxies within the same projected radius as the final assigned
members, but in a velocity range outside the final window used for the
dispersion calculation. Examination of Fig.~\ref{velhists} shows that
there are only a few galaxies outside the final velocity windows used
for Clusters A and D, but less than $\Delta v=4000$ km s$^{-1}$ away.
This provides an estimate of the density of interlopers in velocity
space near each cluster. As an example, consider Cluster A, using the
1.0 $h^{-1}_{70}$ Mpc radius. The final velocity window containing all
members used to calculate the dispersion is a top hat with half-width
$dv=1493$ km s$^{-1}$. We count all galaxies $N_{int}$ in the same 1.0
$h^{-1}_{70}$ Mpc radius, but with velocities between $-2dv...-dv$ and
$dv...2dv$. If the background galaxy distribution is constant with
velocity, $N_{int}$ is also the expected number of interlopers within
the cluster velocity range ($-dv$ to $+dv$). For Cluster A, we find
$N_{int}=(1,3,3)$ for R$=0.5, 1.0, 1.5 h^{-1}_{70}$ Mpc radii. If
these are all taken to be interlopers, they correspond to possible
contamination rates of 5\%, 9\% and 8\%, respectively. For Cluster D,
we find similar results with $N_{int}=(1,2,6)$ and contamination rates
of 3\%, 4\% and 8\%. This analysis assumes that {\em all} galaxies
with velocities between $-2dv...-dv$ and $dv...2dv$ are interlopers,
thus providing an estimate of the maximum effect that they could have.

To see if the presence of such objects could effect the velocity
dispersions, we perform a Monte Carlo simulation with Clusters A and
D, using all three projected radii. We remove $N_{int}$ galaxies from
the velocity range $-dv...dv$. One galaxy is removed in each velocity
bin of width $2dv/N_{int}$, ensuring that the likelihood of removal is
flat across the entire velocity range, corresponding to contamination
by a constant background. The velocity dispersion and associated
errors are recomputed using the new sample, and this procedure is
repeated 1000 times for each cluster and radius combination. We
compute the mean velocity dispersion (using the biweight dispersion
estimator from ROSTAT) and mean error (jackknife of the biweight from
ROSTAT) for each cluster and radius combination from the 1000 Monte
Carlo runs. We then examine (a) the difference between these velocity
dispersions and the original estimates and (b) the rms scatter within
the 1000 Monte Carlo runs. We find that the velocity dispersions
calculated after removing potential interlopers are only $1-5\%$
lower than the initial estimates. As an example, the original
dispersions for Cluster A using the three radii are 532, 619, and 682
km s$^{-1}$. The interloper-corrected estimates are 532, 595, and 647
km s$^{-1}$, corresponding to reductions of 0\%, 4.0\% and 5.4\%,
respectively. The scatter among the Monte Carlo runs is 10-50 km
s$^{-1}$, depending on the cluster and radius used. Adding this in
quadrature to the already quoted error estimates would only increase
the errors by $\sim10\%$. These results demonstrate that there is
little bias in our velocity dispersions due to interlopers.

\subsubsection{Dependence of Velocity Dispersions on Sampling and Galaxy Colors}

In addition to the possibility of interlopers, the spectroscopic
sampling is significantly variable from cluster to cluster, as
described in \S3. This is especially true for Clusters A and D, which
have extensive LRIS data where no color cuts were applied. Because the
measured velocity dispersions for these (and other) clusters in Cl1604
have decreased significantly compared to our and others' earlier
observations \citep[][P98, P01]{gal04}, we examine how the sampling
(by instrument and by color) affects the current results.

First, for Clusters A and D, we simply excluded the LRIS redshifts
and measured their velocity dispersions using only the DEIMOS
data. The DEIMOS data have significantly smaller redshift errors, and
using only the DEIMOS data makes the instrumental sampling similar for
all of the clusters. We use the 1 $h_{70}^{-1}$ Mpc radius for this
test to maintain a significant number of galaxies. Cluster A has 26
galaxies with Q$>2$ DEIMOS redshifts, yielding a velocity dispersion
$\sigma=691\pm103$ km/s, consistent with the results
including LRIS (32 galaxies, $619\pm96$ km/s). For Cluster D, the
DEIMOS-only dispersion is $\sigma=445\pm134$ km/s from 29 galaxies,
compared to $\sigma=590\pm112$ km/s from 53 galaxies when LRIS data
are included. These values are also consistent within the errors,
although the lowered dispersion when only DEIMOS data are used may be
due to the large number of LRIS redshifts excluded. The LRIS-observed
galaxies are, on average, bluer than the DEIMOS-observed ones, and may
therefore have a different dispersion. We note, however, that nearly
half the DEIMOS targets are also outside the red sequence. We conclude
that the inclusion of LRIS data for some clusters does not
significantly alter our results.

We further examined if the computed dispersion is different for red
sequence galaxies than for the cluster population as a whole.  Such
segregation was already shown by \citet{zab93}, who found higher
dispersions for late-type galaxies than for early-types in rich, low
redshift clusters. This effect should be even more pronounced at high
redshift, where the clusters are less evolved. Furthermore, since
there is a color-density relationship, using only the red galaxies
should reduce the contribution of infalling groups/filaments at the
cost of smaller sample size. We applied the same color limits used for the
density mapping ($1.0\le(r'-i')\le1.4$, $0.6\le(i'-z')\le1.0$) and a
magnitude limit of $i'\le 24.0$ to the spectroscopic samples in each
cluster. Only Clusters A, B and D have
sufficient (though still small) numbers of red galaxies for this
exercise. Using the 1 $h_{70}^{-1}$ Mpc radius to maintain a significant number of galaxies, we find: \\

\noindent Cluster A: full sample 32 galaxies, $\sigma=619\pm96$ km/s ; color cut 18 galaxies $\sigma=327\pm62$ km/s \\
\noindent Cluster B: full sample 32 galaxies, $\sigma=811\pm76$ km/s ; color cut 12 galaxies $\sigma=778\pm127$ km/s \\ 
\noindent Cluster D: full sample 53 galaxies, $\sigma=590\pm112$ km/s ; color cut 16 galaxies $\sigma=338\pm71$ km/s \\

In all three cases, the red galaxies have a lower velocity
dispersion. For Cluster A the difference is $3.0\sigma$, while for
Cluster D it is only $2.2\sigma$, and less than $1\sigma$ for Cluster
B.  As shown above, these changes are not the result of instrumental
coverage (DEIMOS or LRIS), so the differences must reflect the
properties of the distinct galaxy populations. It is interesting to
note that our Chandra observations find Cluster A to be the most X-ray
luminous system in the supercluster, exhibiting a relaxed and
well-established intra-cluster medium \citep{koc08}. If the system
formed at an earlier epoch than Clusters B and D, it is expected that
the primordial red galaxy population would have had more time to fully
virialize and establish a much different dispersion than any infalling
blue galaxy population.  The lack of a significant difference in the
velocity dispersions of blue and red galaxies in Cluster B, and to a
lesser extent Cluster D, may be further evidence that the systems are
undergoing collapse or possible merger processes. These results again
suggest that one must exercise caution both when selecting galaxies for
spectroscopic followup using colors and when interpreting the
resulting cluster velocity dispersions. Our sample is extremely
limited, but is one of the first to have so many cluster members at
redshifts near unity. The fact that the pattern of velocity dispersion
as a function of galaxy color is not the same from cluster to cluster
may imply that followup of only red-sequence galaxies might yield
biased or inconsistent results, especially if the red vs. blue galaxy
dispersion depends on the cluster mass or formation epoch. This could
also be true if only the most luminous cluster galaxies are used since
they are much more likely to have red colors. Larger samples with deep
spectroscopy of clusters in various evolutionary stages will be
necessary to clarify these findings.

The velocity dispersions computed for Clusters A and D, using all
galaxies within 0.5 or 1 $h^{-1}_{70}$ Mpc radii, are a factor of two
below the initial estimates from \citet{pos98} and \citet{pos01}. The
$\sigma$ values derived here would place these clusters closer to the $L_X -
\sigma$ relation, alleviating the X-ray under-luminosity discussed in
\citet{lub04}. This consistency makes it tempting to assume that these
are the correct dispersions to use; however, the morphological and/or
color sampling is not the same from cluster to cluster (both due to
sampling differences and population variations), and the X-ray
luminosities may also be unreliable due to point source
contamination. These results clearly require detailed comparison to
simulations to see what population mix is best for producing reliable
velocity dispersions that trace the gravitating halo mass. Simplistic
approaches to measuring cluster dispersions, especially at an epoch
when they are still accreting significant portions of their final
galaxy (and mass) content, are not likely to be reliable except for
the most massive evolved systems.

\subsubsection{Substructure}
\begin{figure}
\plotone{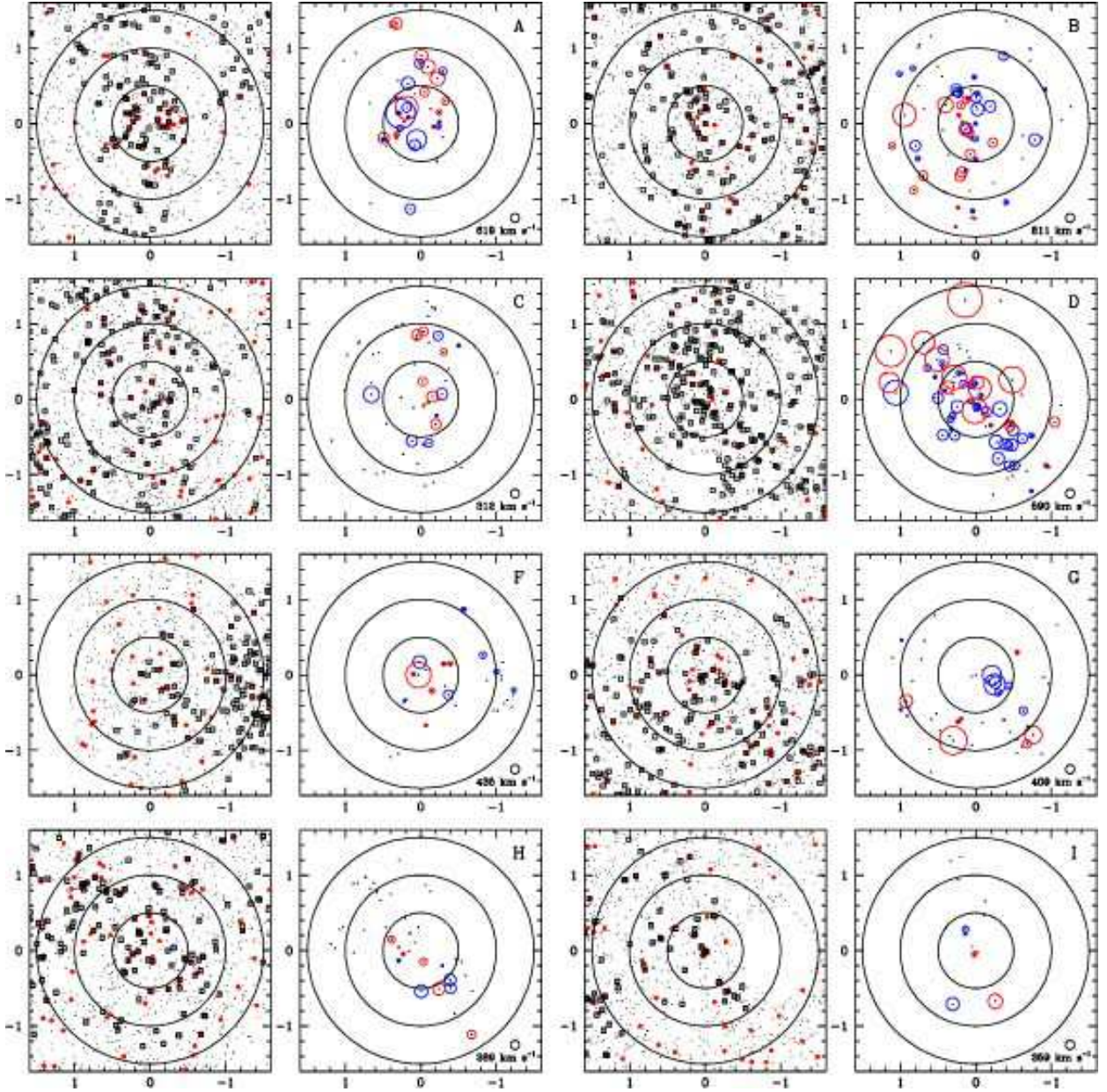}
\caption{Spectroscopic coverage and position-velocity diagrams for 8
clusters in Cl1604. There are two panels for each cluster, each
covering an area of 3.2 $h^{-1}_{70}$ Mpc on a side. The left panels
show all galaxies with $20.5<i'<24$ as small black dots, while red
galaxies used to make the density map are shown as larger red
dots. Black squares outline the spectroscopic targets. The right
panels for each cluster plot all galaxies in the supercluster redshift
range ($0.84\le z \le 0.96$) as small black dots. The final members of
each cluster are then circled, with red circles for galaxies with
higher recession velocities than the cluster mean, and blue for those
galaxies with lower velocities. The circles are scaled by
$v_{gal}/\sigma_{clus}$, with the scaling shown in the lower right
corner of each panel. In both panels, the three large circles
correspond to radii of 0.5, 1.0 and 1.5 $h^{-1}_{70}$ Mpc, used to
measure velocity dispersions.}
\label{posvel} 
\end{figure}

To better understand the individual cluster structures and possible
contamination from neighboring clusters and filaments, we examined the
spectroscopic coverage and position-velocity distributions within each
cluster.  These are shown in Figure~\ref{posvel} for all clusters
except E (which has no clear redshift peak) and J (which has no
spectroscopic coverage). There are two panels for each cluster, each
covering an area of 3.2 $h^{-1}_{70}$ Mpc on a side. The left panels
show all galaxies with $20.5<i'<24$ as small black dots. Essentially,
these are all possible spectroscopic targets in the field regardless
of priority. Red galaxies used to make the density map are shown as
larger red dots. The concentrations in the centers of most clusters
are evident, and in later DEIMOS masks these were prioritized. Black
squares outline the galaxies with spectroscopic
redshifts. Spectroscopic targets which met the density map color
criteria are almost always cluster members, highlighting the efficiency of
our color cuts in selecting supercluster members. Clusters A and D have
very dense coverage because they were the first to be discovered by
\citet{gho86} and were the targets of initial LRIS spectroscopy. The
three large circles correspond to radii of 0.5, 1.0 and 1.5
$h^{-1}_{70}$ Mpc, used to measure velocity dispersions. The right
panels for each cluster plot all galaxies in the supercluster redshift
range ($0.84\le z \le 0.96$) as small black dots. The final members of
each cluster (as determined using ROSTAT within the 1.5 $h^{-1}_{70}$
Mpc radius) are then circled. Red circles are used for galaxies with
higher recession velocities than the cluster mean, while blue circles
show those blueshifted relative to the cluster mean. The circle
sizes are scaled by the ratio of the galaxies' cluster-centric radial
velocity to the cluster's velocity dispersion,
$v_{gal}/\sigma_{clus}$. The corresponding velocity dispersion is
shown in the lower right corner of each panel, along with a circle for
$v_{gal} = \sigma_{clus}$. Cluster E has no redshift limits marked and
no velocity dispersions measured due to contamination from B and C.

Examination of Fig.~\ref{posvel} demonstrates that even the very
well-sampled clusters (A, B and D), with over thirty members within a
1 $h^{-1}_{70}$ Mpc radius, show visual evidence for
substructure. Cluster D in particular appears very elongated in the
NE-SW direction. Cluster B shows velocity segregation, which could be
interpreted as either substructure or a triaxial cluster, elongated in
the radial direction and oriented at a slight angle to the
line-of-sight. To quantify the amount of substructure, we performed
\citet{dre88} tests on Clusters A, B and D using the spectroscopically
confirmed members within a 1.0 $h^{-1}_{70}$ Mpc radius (32, 32 and 53
galaxies, respectively). Also known as the $\Delta$ or DS test
\citep{pin96}, this statistic looks for velocity substructure among
galaxies near each other in projection as an indicator of merging
components. The DS statistic was shown by \citet{pin96} to be the most
sensitive of the five different substructure estimators they examined,
and works well even for moderate samples with only $\sim30$ velocities
per cluster. This test indicates no substructure in Cluster A,
consistent with its appearance in Fig.~\ref{posvel}. Cluster B,
although showing two velocity subclumps (Fig.~\ref{velhists}), has no
significant substructure based on the DS test. This may be the result
of having two subclumps very well aligned along the line-of-sight, a
situation to which the DS-test is not sensitive.  Cluster D has the
highest likelihood of substructure, with the null hypothesis (no
substructure) rejected at the 93\% confidence level. This is consistent
with the evident elongation at an angle to the line of sight and the
velocity segregation between the NE and SE parts of the cluster.

Recent work using both optical and X-ray data
\citep{ser06,paz06,pli06,def05} has shown that many clusters and
groups are quite strongly triaxial, consistent with large simulations
such as those of \citet{jin02}. Additionally, the Cl1604 supercluster
is extremely elongated along the line-of-sight, and we expect, based
on Hubble volume simulations, that the clusters will be aligned with
the supercluster \citep{lee07}. Further understanding of the cluster
dynamics will require detailed comparison to large $N$-body
simulations as well as modeling of the Cl1604 supercluster in
particular.

\subsubsection{Color-magnitude diagrams}

The presence of a strong color-magnitude relation (CMR), the red
sequence for early-type galaxies, has long been observed
\citep{vis77,bow92} and used for cluster detection \citep{gla00}. At
the redshift of the Cl1604 supercluster, galaxies become too small to
classify morphologically from ground-based images. Recent work has
used HST imaging to examine the CMR in high-redshift clusters
\citep{for03}, and in particular Cl1604+4304 and Cl1604+4321
\citep{hom06}. In a forthcoming study, we will combine our extensive
spectroscopic database with a new HST ACS mosaic of the Cl1604
supercluster to extend such work to a wider range of
environments. Here, we briefly discuss the ground-based CMDs of the
individual clusters.

\begin{figure}
\plottwo{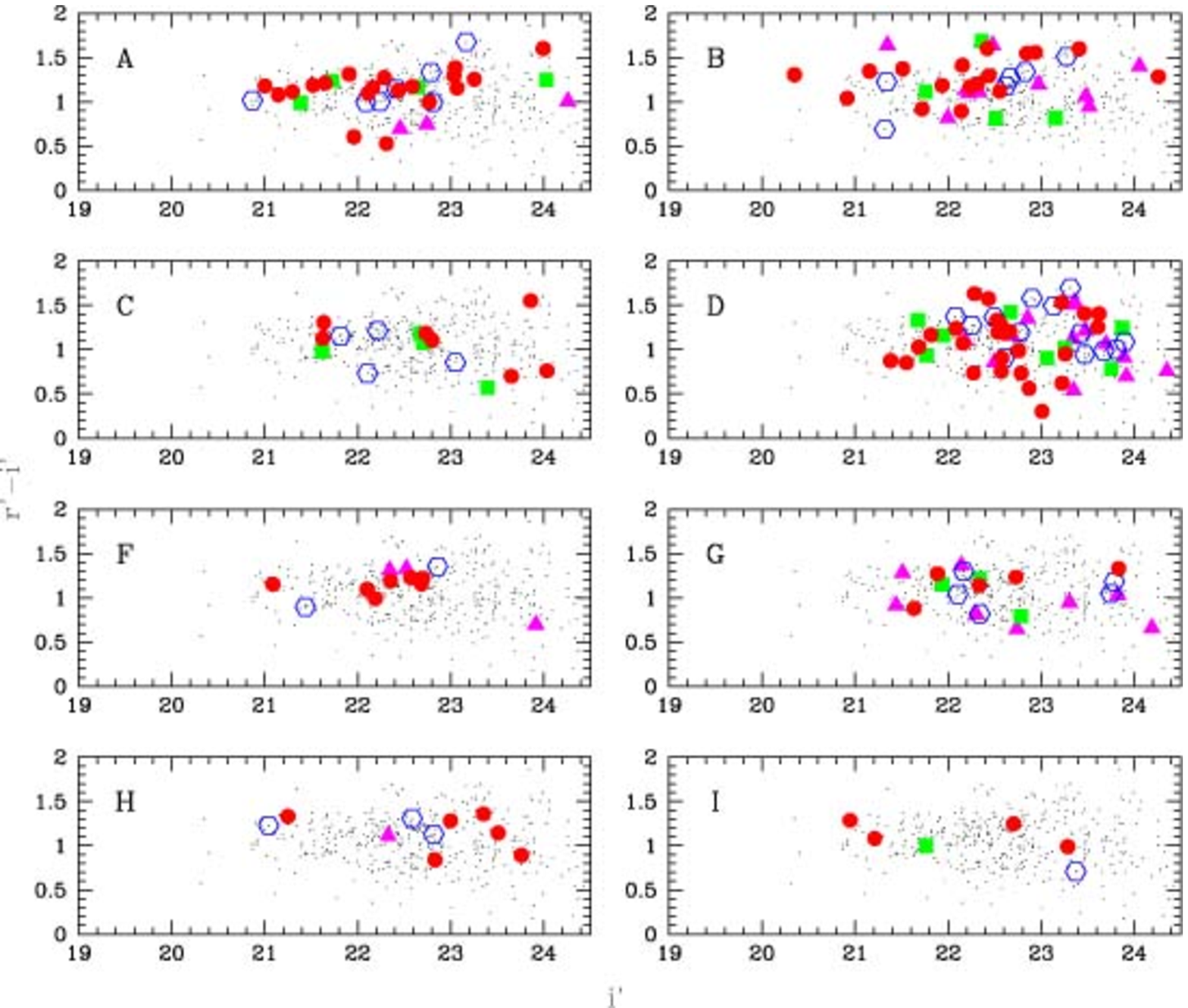}{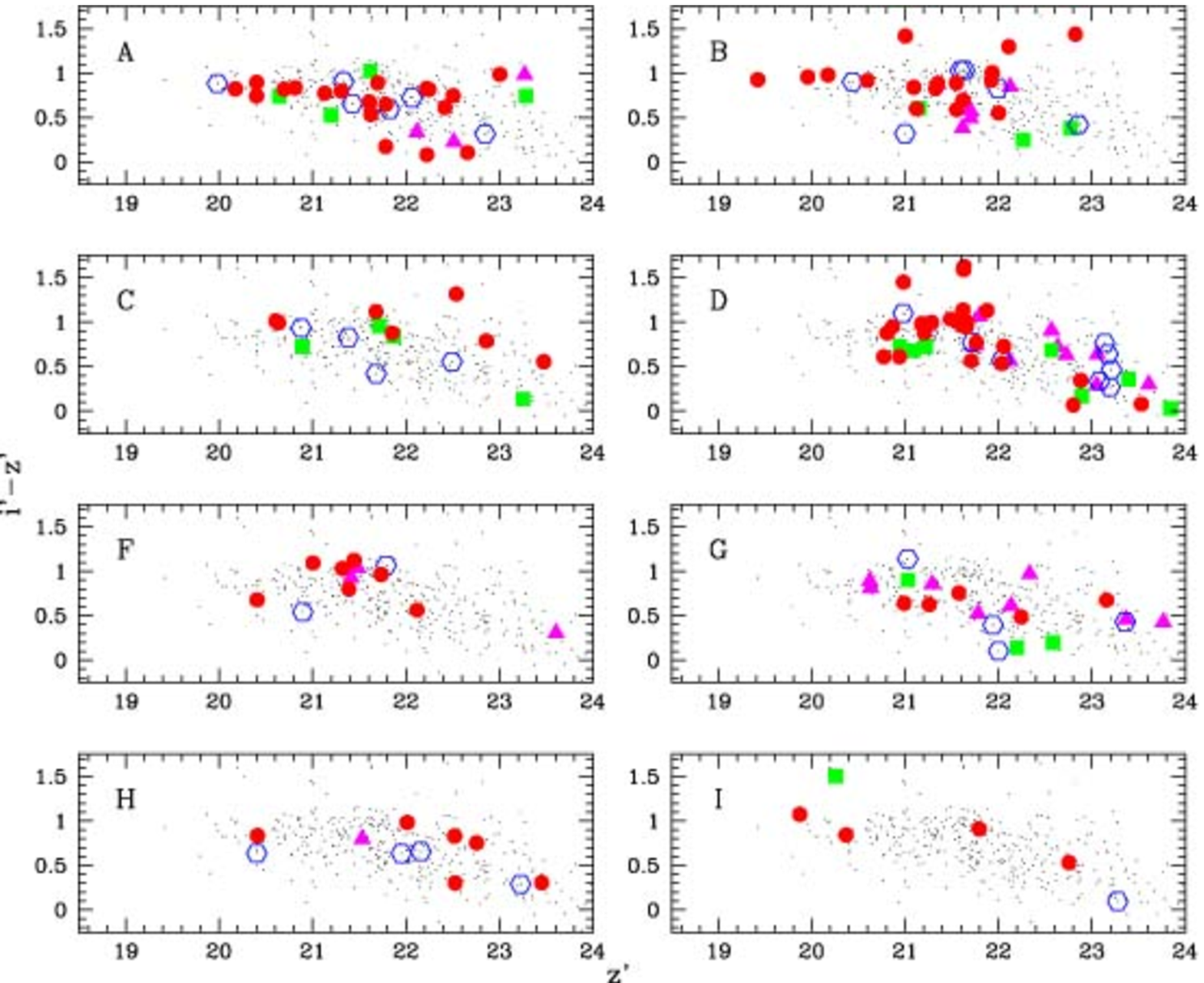}
\caption{The $(r'-i')$ vs. $i'$
(left) and $(i'-z')$ vs. $z'$ (right) CMDs for eight of the clusters in
Cl1604. Small black points indicate all supercluster members.  Large red dots are galaxies determined to be members of
each specific cluster at radii $r < 0.5 h^{-1}_{70}$ Mpc. We plot open blue hexagons for galaxies at $0.5 < r < 0.75
h^{-1}_{70}$ Mpc, filled green squares for $0.75 < r < 1.0 h^{-1}_{70}$ Mpc, and filled magenta triangles  for $1.0 < r < 1.5 h^{-1}_{70}$ Mpc from the cluster center. }
\label{cmrs} 
\end{figure}

Figure~\ref{cmrs} shows the $(r'-i')$ vs. $i'$ and $(i'-z')$
vs. $z'$ CMDs for each of the clusters (note that E is
excluded; see \S4.1.1) in Cl1604 using only objects determined to be
photometrically clean in the ground-based LFC and COSMIC
imaging. Small black points indicate all galaxies in the full imaged
area having redshifts consistent with supercluster membership
($0.84\le z \le 0.96$). Large red dots are galaxies determined to be
members of each specific cluster at radii $r < 0.5 h^{-1}_{70}$
Mpc. We plot open hexagons for galaxies at $0.5 < r < 0.75
h^{-1}_{70}$ Mpc, filled squares for $0.75 < r < 1.0 h^{-1}_{70}$ Mpc,
and filled triangles for $1.0 < r < 1.5 h^{-1}_{70}$ Mpc from the
cluster center.

First, we note that the photometric errors are significant for fainter
galaxies. At $i'\sim24$, for instance, the typical error is 0.1$^{\rm mag}$;
added in quadrature with another band, we expect at least 0.15$^{\rm mag}$ of
scatter at $i'\sim24$ from photometric uncertainties alone. The
intrinsic width of the red sequence, due to age differences at fixed
galaxy mass, is only $\sim0.05^{\rm mag}$ in clusters at $z\sim1$
\citep{hom06,mei06}, similar to that observed locally
\citep{mci05}. The expected slope of the red sequence, a consequence
of metallicity changes with galaxy mass, is also small, of order
$-0.05$ \citep{kod97}. Thus, our ground based observations are not
sufficiently accurate to measure these quantities.

\begin{figure}
\plotone{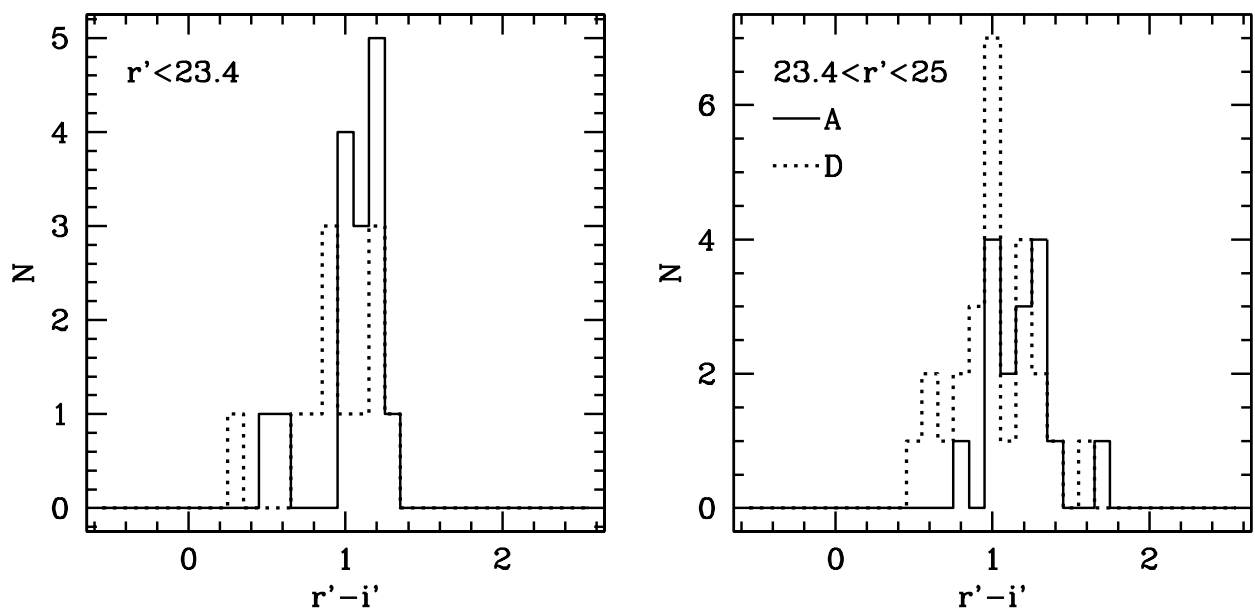}
\caption{The distribution of $r'-i'$
colors for spectroscopically confirmed members of Clusters A and D. The left panel shows galaxies
with $r'\le23.4$, the depth of the complete LRIS spectroscopy in A and
D. The right panel shows fainter galaxies, with $23.4<r'<25$. The
solid histograms indicate galaxies in A, while the dotted line is for
Cluster D. The excess of bluer [$(r'-i')\sim0.9$] galaxies and lack of luminous red galaxies in Cluster
D is evident.}
\label{colordist} 
\end{figure}

Nevertheless, a few general observations are possible. The red
sequence is most prominent in Clusters A and B, which also have the
highest velocity dispersions, consistent with the morphology-density
and color-density relations \citep{dre80,smi05,nui05,coo07} and its
disappearance in low density environments by $z\sim1$ \citep{tan04}.
Cluster D, with a velocity dispersion only 5\% lower than Cluster A,
shows a distinctly less pronounced red sequence. This is illustrated
in Figure~\ref{colordist}, which shows the distribution of $r'-i'$
colors for spectroscopically confirmed members of Clusters A and
D. The left panel shows galaxies with $r'\le23.4$, the depth of the
complete LRIS spectroscopy in A and D. At these magnitudes, red
galaxies have $i'\le22.6$, with photometric errors of $\sim 0.04^{\rm
mag}$ in $r'$ and $\sim0.03^{\rm mag}$ in $i'$, or color errors of
$\sim0.05^{\rm mag}$.  The right panel shows fainter galaxies, with
$23.4<r'<25$, where the errors are much larger. The solid histograms
indicate galaxies in A, while the dotted line is for Cluster D. The
excess of bluer ($(r'-i')\sim0.9$) galaxies in Cluster D is evident,
especially in the fainter population.

Looking at the left panel, Cluster A has eleven
luminous red ($1.0\le r'-i' \le1.4$) galaxies, while Cluster D
(Cl1604+4321), although apparently rich, has only six. If we examine
the most luminous objects, the brightest red galaxy in D has
$r'=22.985$; Cluster A has five more luminous galaxies, extending to
$r'=22.18$. The lack of luminous red galaxies coupled with the
presence of numerous bluer galaxies in D results in a dilution of the
red-sequence, making it appear much less prominent than in Cluster
A. Clusters A and D have magnitude-limited spectroscopy
for all objects with $R<23.5$ from the \citet{oke98} LRIS fields,
covering a $6'\times 8'$ area ($2.8\times 3.7 h_{70}^{-1}$ Mpc) around
each cluster, so the differences in the luminous galaxy populations
between A and D are certainly real and not selection effects.

The absence of luminous galaxies in D is also not an effect of simply
having fewer galaxies as a result of its lower velocity dispersion
(mass); Cluster A has 1.8 times as many objects with $r'<23.4$ and
$1.0\le (r'-i') \le 1.4$, while we would expect only $\sim10\%$ more
if we assume $M\propto\sigma^2$ and $N_{gals}\propto M$. The
discrepancy only gets worse if we look at the most luminous galaxies
($r'\le23.0$), where there are five times more in A than in D. To
explain this as purely a result of A's large mass, we would have to
increase the dispersion of A by $2\sigma$ while at the same time
decreasing D's dispersion by $2\sigma$, resulting in a mass ratio of
4.8, comparable to, but still lower than, the factor of five
difference in luminous red galaxies.

Similarly, the fainter galaxy population differences between A and D
are also physical, as they are sampled by both second-epoch LRIS and
later DEIMOS slitmasks, with similar coverage. Our findings are
consistent with the HST results of \citet{hom06}, who had fewer
redshifts but more precise photometry.  Combined with its elongated
morphology, these data suggest that Cl1604+4321 (D) is a cluster in the
process of formation, and therefore has not yet had time to build up
its red sequence and assemble massive red galaxies. The remaining
structures have insufficient spectral coverage or too few members with
clean ground-based imaging to make conclusive statements about their
galaxy populations.

However, it is clear that many galaxies in the supercluster are
associated {\em not} with the individual clusters but with the
connecting filaments instead. This implies that studies of the cluster
CMR at high redshift that rely on purely photometric data may
overestimate the width of the red sequence due to contamination by
dynamically dissociated galaxies. Estimates of the blue galaxy
fraction may be similarly compromised. Studies of low-redshift
superclusters have also found that $\sim30\%$ of supercluster members
are not associated with specific clusters \citep{sma98}; this fraction
will only increase at larger lookback times as the structures are less collapsed.

\subsection{Three Dimensional Supercluster Structure}

As noted above, many galaxies in the redshift range of the
supercluster are not associated with any of the clusters/groups. We
use our entire redshift catalog of 413 Q$>2$ confirmed members to construct
a three-dimensional map of the supercluster. To produce this map, we
assume that individual galaxy redshifts directly reflect the position
of the galaxy within the structure. This is not strictly correct,
since the clusters within the structure can have significant
velocities relative to the Hubble flow, and the galaxies within the
clusters also have their own peculiar velocities. However, we have no other
data to constrain the dynamics within this region. If the structure is
bound, then the clusters themselves may have significant peculiar
velocities within the supercluster, further complicating any dynamical
analysis. The redshift depth of the supercluster, taken at face value,
implies a radial length of order $100~h_{70}^{-1}~{\rm Mpc}$, while
the transverse size (using Clusters A and J) is only
$\sim13~h_{70}^{-1}~{\rm Mpc}$.

\begin{figure}
\plotone{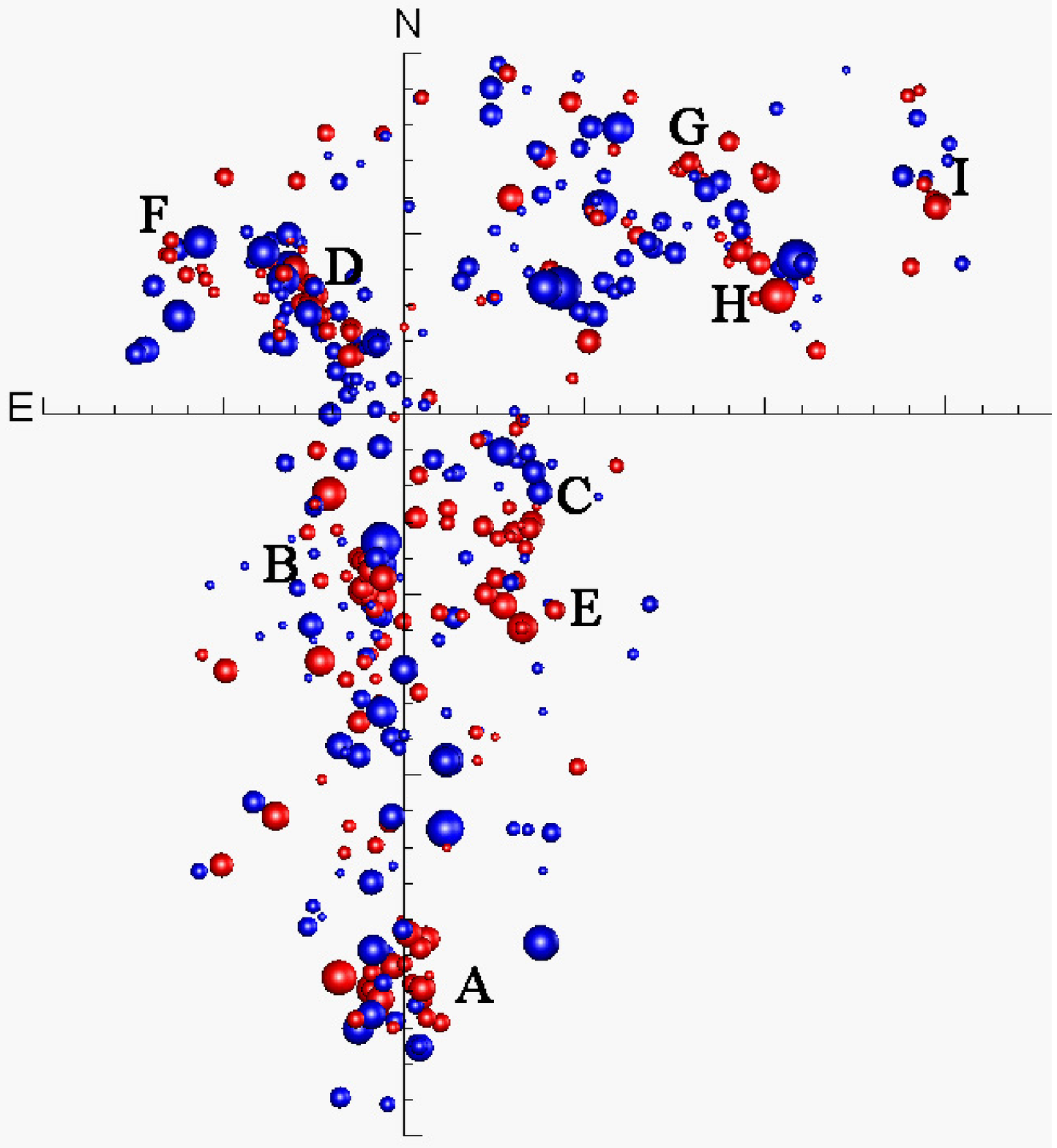}
\caption{Face-on (xy) view of the Cl1604 supercluster. We show red and blue galaxies, divided at $(i'-z')=0.7$, as correspondingly colored spheres. Each sphere is scaled by the observed $i'$ luminosity of the galaxy. 
}
\label{threed1} 
\end{figure}

\begin{figure}
\plotone{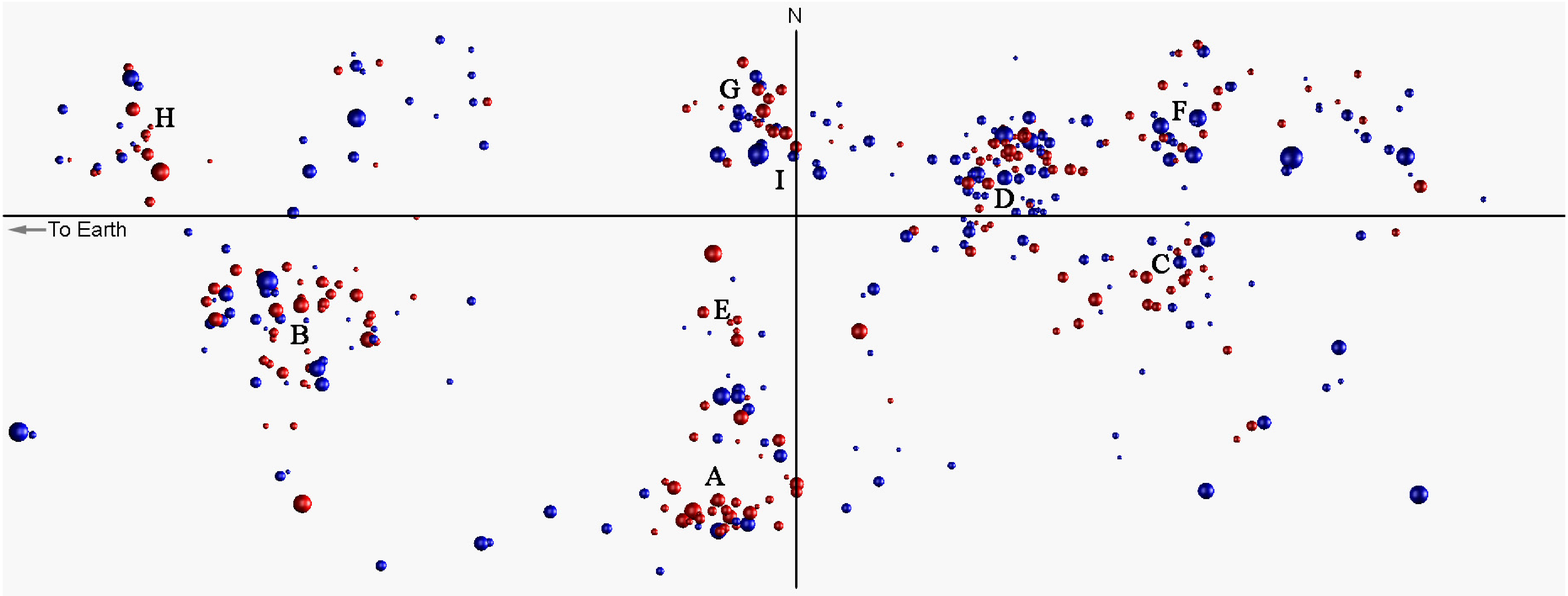}
\caption{Radial (yz) view of the Cl1604 supercluster. We show red and blue galaxies, divided at $(i'-z')=0.7$ as correspondingly colored spheres. Each sphere is scaled by the observed $i'$ luminosity of the galaxy. 
}
\label{threed2} 
\end{figure}

\begin{figure}
\plotone{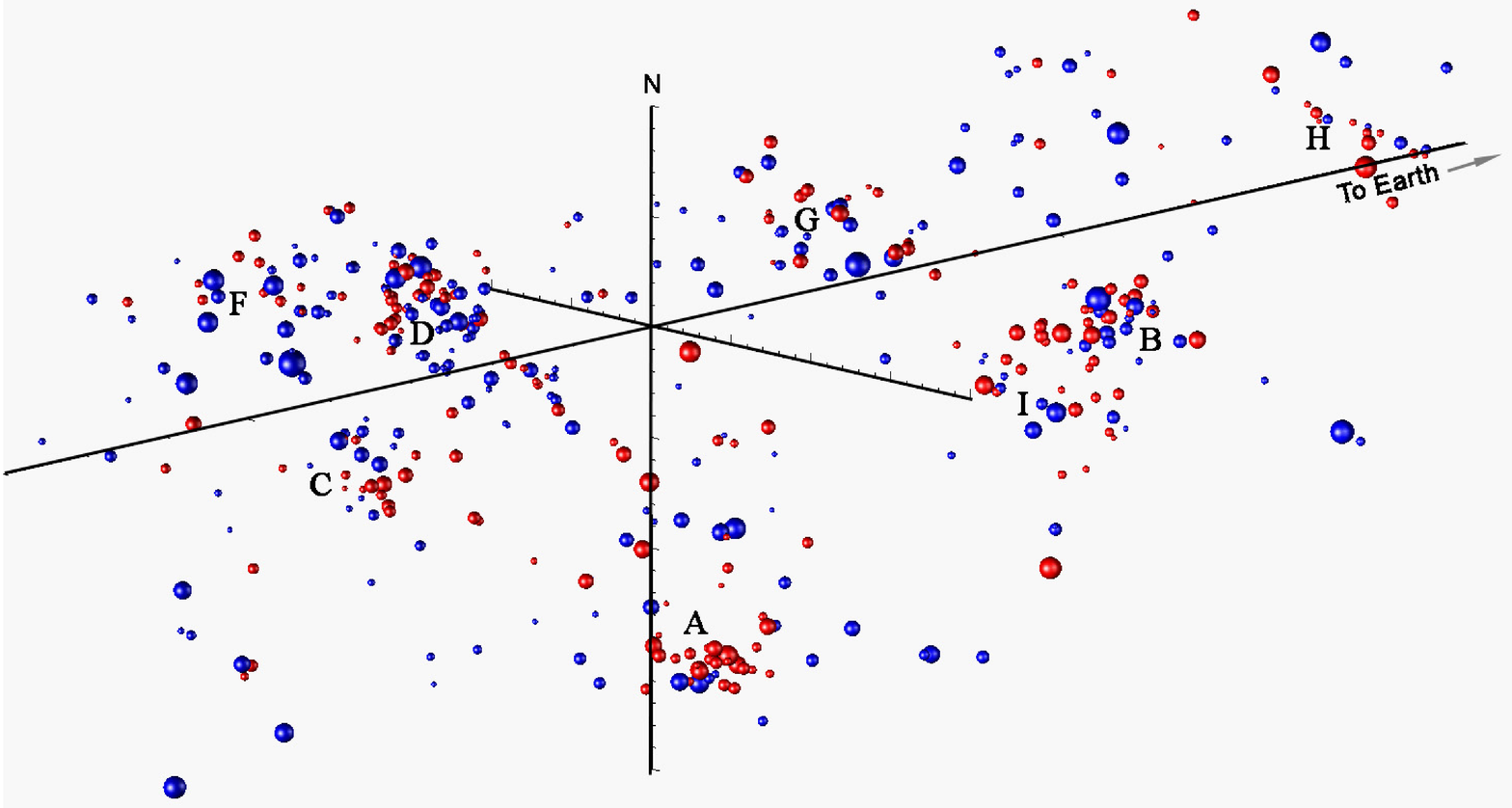}
\caption{Rotated view of the Cl1604 supercluster. We show red and blue galaxies, divided at $(i'-z')=0.7$ as correspondingly colored spheres. Each sphere is scaled by the observed $i'$ luminosity of the galaxy. 
}
\label{threed3} 
\end{figure}

Figures~\ref{threed1}, ~\ref{threed2} and ~\ref{threed3} show three
views of the supercluster. Because of the nearly 8:1 axis ratio, we
have compressed the radial axis by a factor of five. We show red and
blue galaxies, divided at $(i'-z')=0.7$ as correspondingly colored
spheres. Each sphere is scaled by the observed $i'$ luminosity of the
galaxy. Figure~\ref{threed1} shows the face-on view of the
supercluster, as observed on the sky. Clusters A, B and D can be
clearly seen. The large red galaxy population in A is evident, as is
the elongation of Cluster D. Figure~\ref{threed2} shows the radial
distribution of galaxies in Cl1604 as a function of declination. The
individual clusters are clear, along with the large number of
intercluster galaxies. Color segregation is easily seen, although
luminous red galaxies do exist far outside the cluster centers, and
luminous blue galaxies are present in the lower-mass clusters and
groups. The most luminous blue galaxies are almost exclusively found
in the filaments, far from the cluster cores. There are also large
voids in the overall galaxy distribution, consistent with the
structures seen in large simulations. Finally, Figure~\ref{threed3}
shows a rotated view of the supercluster, making its overall structure
clearer. From these figures it appears that Cluster B at $z=0.8656$
is relatively isolated from the rest of the supercluster. Determining
whether it is actually bound to the overall structure or a more
isolated foreground object will require detailed $N$-body simulations
mimicking this supercluster.

\section{Fore- and Background Structures}

Besides the supercluster, a variety of apparent fore- and background
structures can be seen in Figure~\ref{spechist}. We apply three
different criteria to identify coherent overdensities in both redshift
and in projection. Candidate structures are selected by taking each
galaxy with a measured redshift as a seed, and searching for peaks
both in redshift and in projection. We look for groups of galaxies
that meet at least one of the following three requirements:
\begin{itemize}
\item A. 15 or more concordant redshifts within $\Delta z=0.01$, no spatial requirement
\item B. 5 or more concordant redshifts within $\Delta z=0.01$ and within a $0.5 h^{-1}_{70}$Mpc radius
\item C. 5 or more concordant redshifts within $\Delta z=0.01$ and within a $1.0 h^{-1}_{70}$ Mpc radius
\end{itemize}

The redshift interval of $\Delta z=0.01$ used in peak selection
corresponds to a velocity width of $\sim 1500$ km s$^{-1}$, broader
than the expected dispersion of all but the most massive clusters, and
is identical to the criterion of \citet{gil07}. Criterion A includes
structures like walls or filaments which might be rejected by the
latter. Criteria B and C more closely match other surveys that
do spectroscopic followup in a limited spatial region around candidate
clusters. Criterion A selects a total of 18
redshift peaks outside the supercluster redshift range. Criterion B
selects 6 peaks, all of which are also found by A, while Criterion C
selects 13 peaks, 11 of which are also selected by A. Although these
candidate structures are chosen based on well-defined criteria, the
uneven spatial coverage, survey depth, color selection, and biases
such as easier identification of emission-line objects imply that this
sample is neither complete nor unbiased (in redshift or galaxy type).

\begin{figure}
\plotone{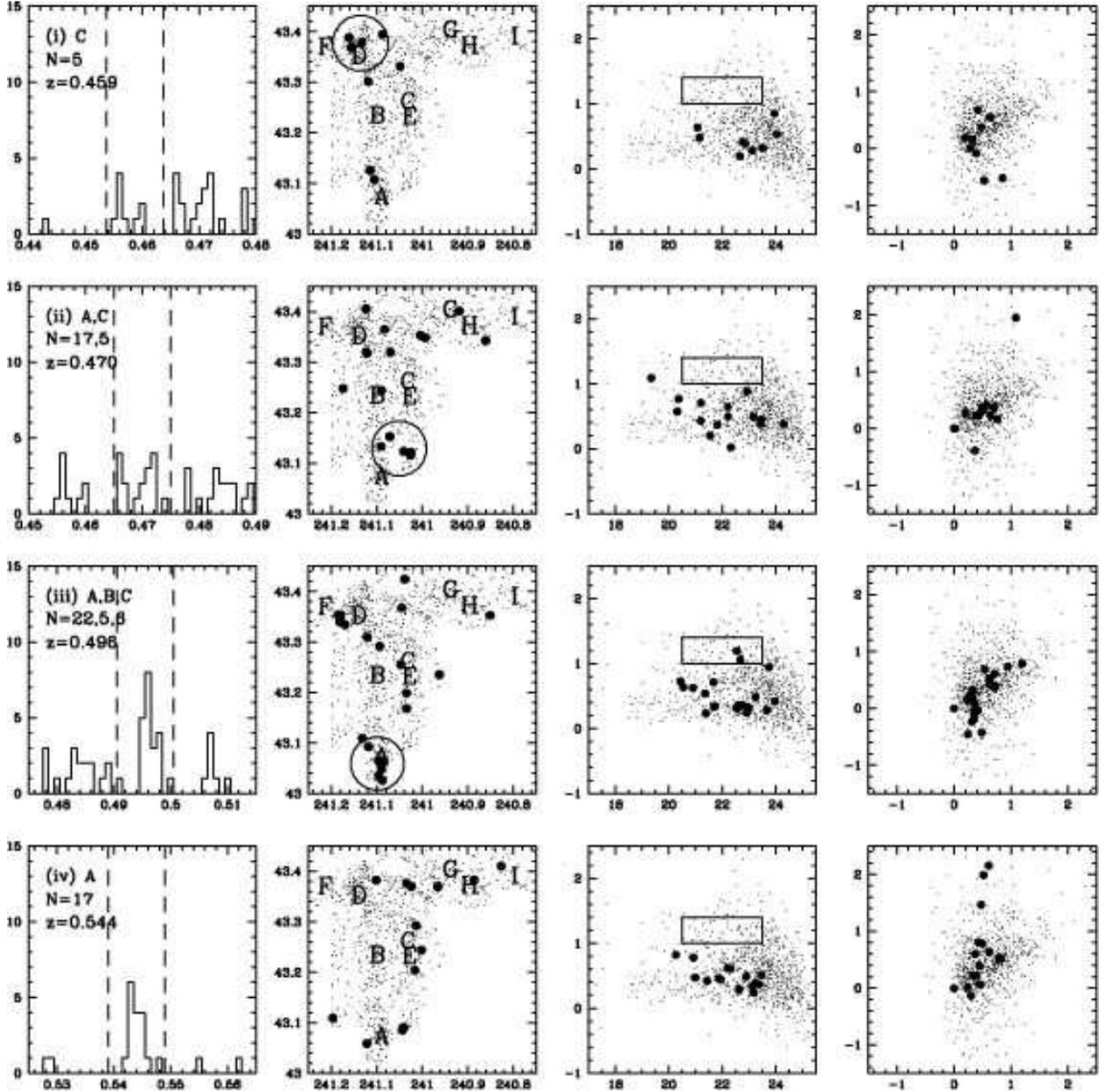}
\caption{The spatial distributions and CMRs of galaxies in each fore-
and background redshift peak. Each row corresponds to a different peak
in the redshift histogram, starting at $z\sim0.4$. The first column
shows the redshift histogram in the region of the specific redshift
peak, with dashed lines demarcating the redshift range used to plot
galaxies in the spatial and color distributions. The left panel
includes a Roman numeral corresponding to the structures tallied in
Table~\ref{losinfo}, the number of galaxies in the peak, and the
redshift. The second column shows the projected distribution of
galaxies in the redshift peak (large dots) overlaid on the overall
distribution of spectroscopic objects. Cl1604 components A-J are
marked. For redshift peaks that appear to be real structures, a circle
of radius 1.5 $h^{-1}_{70}$ Mpc is drawn around the structure
centroid. The third column presents the $r'-i'$ vs. $i'$ CMD with all
photometric objects as small black dots and the members of each
structure as large red dots. A black rectangle outlines the color and
magnitude range used in making the density map. The fourth column
shows the $r'-i'$ vs. $i'-z'$ color-color diagram using the same
symbols.  }
\label{bgclus} 
\end{figure}

Whether or not these structures are real or artifacts of our color
selection and spatial sampling requires looking in greater detail at
the spatial distributions and CMRs of galaxies in each apparent
redshift peak. These are plotted in Figures~\ref{bgclus}a-e. Each row
corresponds to a different peak in the redshift histogram, starting at
$z\sim0.4$. The first column shows the redshift histogram in the
region of the specific redshift peak, with dashed lines demarcating
the initial redshift range $\Delta z=0.01$ used to select the peak
with Criterion A. The left panel includes a Roman numeral
corresponding to the structures tallied in Table~\ref{losinfo},
followed by letters denoting which criteria are met by the peak. We
also show the number of galaxies in the peak based on each of the
criteria by which it is selected. The median redshift is included,
again using all Criterion A galaxies. The second column shows the
projected distribution of galaxies in the redshift peak (within
$\Delta z=0.01$) as large dots overlaid on the overall distribution of
spectroscopic objects. The locations of Clusters A-I are labeled. If
there is an apparent projected overdensity of objects as defined by
Criteria B and/or C, we plot a circle of radius $1.0 h^{-1}_{70}$ Mpc
around the structure centroid, using the mean positions of the
galaxies within the apparent projected overdensity (except for
structure vii, as detailed below). When there are multiple spatial
overdensities in a single redshift peak meeting Criteria B and/or C,
we show only the one with the most members. The third column presents
the $(r'-i')$ vs. $i'$ color-magnitude diagram with all photometric
objects as small black dots and the structure members (based on
Criterion A) as large dots. The color and magnitude limits for
galaxies used in making the density map are shown as the black
rectangle in these panels; the highest priority DEIMOS targets have
these colors but extended to $i'=24$. The fourth column shows the
$(r'-i')$ vs. $(i'-z')$ color-color diagram using the same
symbols. Redshifts and positions are shown for all spectroscopic
objects, while photometric data points are shown only for those
objects with clean ground-based photometry to avoid incorrect colors.

Table~\ref{losinfo} provides information for all twenty of these
redshift peaks. Column 1 gives the Roman numeral identifier of the
peak. Columns 2 and 3 give the J2000.0 coordinates. For peaks selected
only by Criterion A, the coordinates are the mean of all galaxies in
the $\Delta z=0.01$ window; because this criterion does not require a
spatial concentration, the coordinates are not necessarily reflective
of any true structure center. For peaks selected by Criteria B and/or
C, we use the mean coordinates from the galaxies within the
overdensity defined by Criterion C, which is most likely to provide a
meaningful position. Column 4 lists which criteria were met by each
peak, while Column 5 gives the median redshift using all Criterion A
galaxies. Columns 6, 7 and 8 give the number of galaxies meeting
Criteria A, B and C, respectively.

Figure~\ref{bgclus} shows a remarkable diversity of structures. Some
are apparently chance projections or collections of galaxies at very
similar redshifts but with no clear physical association. These types
of objects are likely to be selected by Criterion A but not by B or C,
and include structures iv, xiii, xvi and xvii. Seven out of
20 redshift peaks meeting any of our criteria, or 35\%, fall into this
category. These could be sparse walls or filaments crossing our field,
or simple statistical fluctuations. As noted earlier, the uneven
spatial sampling, sparse coverage in some areas, and changes in
targeting priority make it difficult to model the likelihood of these
being real structures without using a full cosmological
simulation. Other structures are more clearly filaments or walls, and
there are distinct clusters/groups as well. We discuss simple
estimates of the false detection rate in Section 5.2.

\subsection{Individual Redshift Peaks}

We briefly describe each of the structures selected by our three
criteria, including their potential effects on the photometric
detection of Cl1604 components.

\begin{enumerate}
\renewcommand{\theenumi}{\roman{enumi}}
\setlength{\itemsep}{0pt}
\item Selected only by Criterion C, there is a concentration of five
galaxies within 1 $h^{-1}_{70}$ Mpc near Cluster D, and only nine
within this redhift peak anywhere in the field. Most of the galaxies
in this peak were observed in the original LRIS survey of
\citet{oke98}. The colors of these six galaxies reveal no clear red
sequence, so this is not likely to be a real structure.
\item Although galaxies in this peak meet both Criteria A and C, there
is no unambiguous compact, physical structure. Examining only the 11
galaxies in the larger subpeak at $0.468<z<0.474$ highlights the small
clump north of Cluster A. Interestingly, most of the galaxies in this
peak are H$\beta$ and [OIII] emitters as seen in DEIMOS spectra.
\item There are two strong concentrations of galaxies associated with
this redshift peak. In the south, there are 8 galaxies, 7 of which
are within a 1 $h^{-1}_{70}$ Mpc radius, most of which have
spectra from the original LRIS survey of \citet{oke98}. These eight
galaxies have $<z>=0.4961$. In the north, there are 6 galaxies within a
1 $h^{-1}_{70}$ Mpc radius near Clusters D and F, mostly with DEIMOS
spectra, and $<z>=0.4967$. The coordinates and dispersion reported in
Table~\ref{losinfo} are derived using only the ten galaxies in the
southern clump.
\item Selected only by Criterion A, these galaxies are scattered
throughout most of the field. There is no identifiable group or
cluster, despite the modest number of galaxies and the well-defined
shape of the peak. Again, almost all of the galaxies in this peak are
H$\beta$ and [OIII] emitters as seen in DEIMOS spectra.
\item Identified by Criterion C, there are 8 galaxies in the
$\Delta z=0.01$ window. The five galaxies meeting Criterion C are
mostly concentrated in a tight structure only $38''$ or 240 $h^{-1}_{70}$ kpc
across.
\item One of the most interesting structures, this appears to be a
filament running almost directly N-S in front of the supercluster. It
has a low, group-like velocity dispersion of $\sigma=328$ km s$^{-1}$,
while showing a significant red sequence. Nearly 80\% of the peak
members have DEIMOS redshifts; the lack of galaxies in the western
part of the mapped field is therefore not likely due to lack of
coverage. We report the median coordinates and dispersion using all 39
members in the $\Delta z=0.01$ window, although there is no clear
center to this structure, and many subclumps that meet Criteria B
and/or C. The reddest members of this structure meet our color
criteria for the density map, likely enhancing the detectability of
Cl1604.
\item These galaxies are isolated in the northeast corner of the
field, and the majority have redshifts from the \citet{oke98}
survey and would not meet our current color criteria. However, the
lack of galaxies at the same redshift near Cluster A, where there is
also heavy LRIS coverage, suggests that this structure is spatially
isolated.
\item A tight clump of five galaxies is located near Cluster G. Four
out of five of these galaxies have colors falling within our red
galaxy criteria for density mapping, enhancing the detectability of
Cluster G. The four galaxies in closest proximity to each other are
within $\Delta z=0.0025$, or $\sim400$ km s$^{-1}$, suggestive of a
small group.
\item Detected by all three criteria, there is a concentration of 11
galaxies within a 1 $h^{-1}_{70}$ Mpc radius, centered near Cluster
D. Almost all are from the \citet{oke98} survey.  These 11 galaxies
span a velocity range of only $\sim600$ km s$^{-1}$ with
$<z>=0.696$. They yield a velocity dispersion of
$\sigma_{biwt}=218\pm49$ km s$^{-1}$, commensurate with being a
small group. Again, some of these galaxies meet the red galaxy color
cut, enhancing the detectability of Cluster D.
\item A similar structure to the previous one. The large concentration
of 16 galaxies with $<z>=0.7284$ near Cluster H contains only objects
with DEIMOS redshifts, many of which have red colors. They have a
velocity dispersion of $\sigma_{biwt}=285\pm88$ km s$^{-1}$,
typical of small groups.
\item The redshift distribution shows multiple peaks within the
$\Delta z=0.01$ selection window, and at least two spatial
concentrations in the spatial distribution. Just south of clusters B
and E we see 17 galaxies with $<z>=0.775$ within a 1.6 $h^{-1}_{70}$
Mpc radius. They have a velocity dispersion of
$\sigma_{biwt}=352\pm82$ km s$^{-1}$, typical of small
groups. In the northwest, between Clusters H and I, there are 8
galaxies in a 1.25 $h^{-1}_{70}$ Mpc radius. This clump has a mean
redshift of $<z>=0.781$. Unsurprisingly, a
majority of these galaxies have red colors that fall within our
selection window, since their redshifts are quite similar to that of
the supercluster.
\item An extremely narrow redshift peak, there may be a small group
near Cluster G, selected by Criteria B and C. Even using all galaxies
in the peak gives a dispersion of only 133 km s$^{-1}$.
\item Selected only by Criterion A, these galaxies are scattered
throughout most of the field. There is a concentration just west of
Cluster D, but insufficient to meet either of the spatial
criteria. Since galaxies at this redshift would have colors strongly
sampled by DEIMOS, and the region near Cluster D has extensive LRIS
spectroscopy as well, it is unlikely that there is a significant
structure associated with this peak.
\item A possible structure near Cluster D. There are 10 galaxies in
this region, with $<z>=0.822$ and $\sigma_{biwt}=220\pm142$ km
s$^{-1}$. Most of the galaxies are red and seem to form a red
sequence. As with the previous structure, the spectroscopic sampling
in this area is very dense, so it is unlikely that we have missed many
of the luminous red members of this structure.
\item A small clump near Cluster A. All of the galaxies in the tight clump
are from the \citet{oke98} survey.
\item Only selected by Criterion A, the redshift and spatial
distributions show no evidence for a group.
\item Only selected by Criterion A, the redshift and spatial
distributions show no evidence for a group, except for a possible
concentration in the northern end of the field.
\item Although meeting both Criteria A and C, there is no obvious
single overdensity, and the redshift distribution is quite broad.
\item Although the redshift distribution is broad, we see a very
compact group north of Cluster C, and this peak meets all three
selection criteria. The 9 galaxies in this clump are contained in a
region of radius $0.4 h^{-1}_{70}$ Mpc. They have $<z>=1.179$ and
$\sigma_{biwt}=289^{+40}_{-92}$ km s$^{-1}$. This structure seems to
be a low mass group.
\item A group or poor cluster at $z=1.207$. The majority (12) of the
galaxies are concentrated in a $1.5\times2.5 h^{-1}_{70}$ Mpc region
in the northeast corner of the observed area, whose center we have
marked on the plot. The velocity dispersion from these 12 galaxies is
$288\pm82$ km s$^{-1}$. All of these galaxies are very faint ($i'>23$)
and are detected spectroscopically from their OII emission alone,
consistent with their broad range in colors. If the velocity
dispersion is reflective of the group mass, this is the highest
redshift group of such low mass known.

\end{enumerate}

From the third column in Fig~\ref{bgclus} we see that many of the structures, especially at
$z\sim0.7$, contain significant numbers of galaxies with colors
consistent with supercluster membership. Some of these show projected
density peaks near components of the supercluster, which likely
contributed to the detectability of low-mass groups and filaments in
the region. Narrowing the color range slightly would not eliminate
their contributions to the density map. The CMDs and redshift
distributions of the structures demonstrate the difficulty of
establishing the physical nature of moderate-mass, high-redshift
systems.

\subsection{Are Such Structures Real Physical Associations ?}

It is extremely common practice to detect clusters with some
photometric technique, and report as few as three redshifts for
confirmation. For instance, \citet{ols05} examined spectroscopy in the
fields of five $z>0.6$ clusters in the ESO Imaging Survey (EIS) Cluster
Candidate Catalog, and found numerous groups in many of the fields,
often with only three concordant redshifts (such as
EISJ0046-2951). They find multiple redshift peaks in some of their
fields, and rely on the alignment of galaxies in each redshift peak
with the location of the candidate derived from the initial imaging to
determine confirmations and assign redshifts.  They do not specify a
minimum area within which these three redshifts had to be found (with
statistical significance tests relying purely on redshift
distributions with no spatial information), nor detail the
spectroscopic completeness (especially as a function of color), making
statistical statements impossible. However, inspection of their
figures shows that these objects were typically within a $\sim2$
arcminute diameter region inside the large spectroscopic field.
Similarly, \citet{gil07} performed follow-up observations of RCS
\citep{gla05} clusters, and often found only 3-5 concordant redshifts
for the candidates.  They follow the arguments of \citet{gil04},
similar to those of \citet{hol99} and \citet{ram00}, to conclude that
finding just three concordant redshifts for early type galaxies in a
field with 5 arcmin radius at $z=0.3$ results in a $\sim99\%$
likelihood of having a true structure. This radius corresponds to
$\sim2$ arcminutes at $z=1$.

To qualitatively assess the reliability of cluster confirmation using
so few concordant redshifts, we examined each of our redshift peaks
which meet only Criterion A, and {\em not} Criteria B or C. There are
7 such peaks comprising $\sim35\%$
of our candidates. This provides a sample of galaxies that are in
redshift peaks but not contained in any obvious spatial structure. We
could simply assume that all of these are projections, implying that
requiring {\em no} spatial coincidence results in at least a $35\%$
false positive rate for cluster confirmation. We caution that even
candidate structures meeting Criteria B or C may be projections (which
would imply an increase in the false positive rate), and that some
structures that only meet Criterion A may be real (decreasing the
false positive rate). As noted earlier, applying our spectroscopic
selection in detail to large cosmological simulations with realistic
galaxy distributions and colors is needed to definitively understand
the nature of the detected structures.

To compare more directly with the studies mentioned above, we must
account for our higher redshift sampling. For instance, \citet{ols05}
have 266 redshifts in 5 fields. They do not provide the area coverage
of their spectroscopy, so we estimate a spectroscopic area coverage of
$\sim150$ sq. arcmin based on their figures, or a redshift density of
$\sim1.75$ per sq. arcmin. This compares to our 1138 redshifts in
$\sim 300$ sq. arcmin, a sampling rate of 3.8 per
sq. arcmin. Excluding the LRIS redshifts, which cover small regions
around Clusters A and D, we have 3.0 redshifts per sq. arcmin. Our
redshift sampling is indeed greater by a factor of $\sim2$. Our area
coverage is also a factor of $\sim2$ greater. However, we require five
times more galaxies in each peak (15 as opposed to 3), approximately
compensating for these factors.  Thus, our estimated 35\% false
positive rate may be indicative of the rate expected in \citet{ols05}.
If instead we reduce our requirement to account purely for the
difference in sampling, we need only $\sim5$ galaxies per peak, but
contained within a region comparable to the spectroscopic fields of
\citet{ols05}, about $5'\times5'$. This yields 22 candidate peaks, and
they are almost identical to those found from our full area and
requiring 15 galaxies in the peak, suggesting that our false positive
rate is reasonable. Comparison to \citet{gil07} is more difficult,
since they do not provide information on the effective area covered in
each field, magnitude limits, or color selection. However, examination
of their Fig. 2 shows numerous cases where there is no obvious
redshift peak, especially for fields that are poorly sampled.  We also
consider how rapidly the number of candidates increases with area for
a fixed number of galaxies per redshift peak. Our original Criterion B
requires 5 concordant galaxies within a $0.5 h^{-1}_{70}$Mpc radius;
the typical area covered is 5 sq. arcmin. Similarly, Criterion C
encompasses $~\sim17$ sq. arcmin on average, while the EIS area is 25
sq. arcmin. We find 6, 13, and 22 candidates using these search areas,
respectively. Visual inspection of Fig.~\ref{bgclus} suggests that
even Criterion C may yield structures that are not obviously physical
associations, and sampling even larger areas and requiring fewer than
five concordant redshifts can only exacerbate this problem.

Alternatively, the redshift peaks in our data showing minimal
spatial structure (i.e., meeting only Criterion A) provide a test sample to
estimate the likelihood of false positives from small projected galaxy
groupings.  To do this, we examine each galaxy in each of the redshift
peaks selected only by Criterion A.  For each galaxy, we count the
number of neighbors within 0.25, 0.5 and 2 $h^{-1}_{70}$ Mpc radii
(31", 62", 4.13' at $z=1$). We then count how many projected groupings
of $N_{near}>3$ galaxies are present for each peak, mimicking some of
the selections used in other surveys.  Using the smallest radius, 3
out of 6 redshift peaks have a group of at least 3 galaxies within the
required area. Increasing the test radius to 0.5 $h^{-1}_{70}$ Mpc
results in all but one of these redshift peaks containing a group of
three or even four nearby galaxies. For the largest test radius, every
redshift peak contains at least three and as many as five distinct
groups, and these groups typically have 6-11 members. These results
suggest that, with our sampling, requiring only 3 galaxies even in a
very small area is likely to produce false structures. At the largest
radius (2 $h^{-1}_{70}$ Mpc) we are covering areas similar to those in
\citet{ols05} and \citet{gil07}. Adjusting for the lower sampling in
\citet{ols05}, nearly all of our Criterion-A-only peaks would have 3
concordant redshifts within a comparable field size. However, only one
of these peaks would have 4 or more concordant redshifts. This simply
reflects the fact that sparse spectroscopic sampling over modest
regions can easily produce spurious redshift peaks.

We note that this analysis includes not just red sequence galaxies,
since in some cases we do not sample the RS at the redshift of the
peak.  This may increase the likelihood of finding chance
projections. If we required instead a small number of concordant
redshifts and that these objects were also along the red sequence, we
would have many fewer candidates and eliminate many potentially
spurious groups. However, most spectroscopic follow-up studies do not
impose such a requirement. In addition, for poor groups, we might not
expect to see a strong red sequence at redshifts like those studied
here, so requiring a red sequence might eliminate otherwise physical
structures. Our analysis also does not consider the alignment of small
redshift peaks with cluster candidates pre-selected using photometric
redshifts or red sequence techniques. The likelihood of finding a
redshift peak with $N\sim3$ galaxies at the location of a cluster
candidate and with redshift comparable to the photometric estimate is
certainly much lower than finding a peak at any redshift along the
line of sight. However, we have shown that in a modest field, it is
easy to find peaks with $N\sim3$ galaxies at almost any redshift, so
using such data to confirm candidate clusters is questionable.

A competing effect is low spectroscopic completeness,
decreasing the likelihood of finding close projections.  If we use
simple criteria of three galaxies within a velocity window of
$\pm3000$ km s$^{-1}$ and a radius of 0.25 $h^{-1}_{70}$ Mpc, every
redshift peak along the line-of-sight would qualify as a physical
structure. While this may be true, it is unlikely. Given the complex
sampling, we would have to replicate our observations on large mock
galaxy catalogs to give correct likelihoods for each structure. This
problem will persist for surveys like ORELSE, and for large sky
surveys such as Pan-STARRS and LSST, where spectroscopic followup of
complete samples of faint, distant galaxies will be difficult.

\section{Discussion}

\subsection{Implications for Lensing and Cosmology}

Detection of high-redshift clusters using photometric techniques can
be affected by line-of-sight projections over radial distances
comparable to the redshift resolution of simple color-based
techniques. \citet{coh07} show that many clusters detected in a
red-sequence type survey at $z\sim1$ are made up of galaxies from
multiple distinct cluster-size dark matter halos. Their work does not
include smaller groups, of which we expect to find many in the infall
regions of high-redshift clusters. They also demonstrate that the
problem becomes substantially worse at $z\sim1$ than it is at
$z\sim0.5$, in part because a fixed pair of filters cannot sample
features such as the 4000\AA~ break over such a broad redshift range.

We find many small groups in both the fore- and background of the
target structure at $z\sim0.9$. Examination of model galaxy tracks
from \citet{bru03} in the $r'-i'$ vs. $i'-z'$ color-color space for
both galaxies with a single burst at $z=3$ and the RCS model (a 0.1
Gyr burst ending at $z=2.5$ with a $\tau=0.1$ Gyr exponential decline
thereafter) shows that colors for galaxies over a moderately broad
redshift range ($z\sim0.8-1.2$) are quite similar, with the
differences in color comparable to ground-based photometric errors
(ORELSE I). Therefore, any survey relying on only two- or three-color
moderate-depth ground-based imaging is subject to this type of
confusion. The surveys will be biased toward detecting projected
systems, the same problem that has plagued lower redshift optical
cluster catalogs \citep{kat96}. Even with large cosmological
simulations, the detailed effects of such projections on cluster
detection, number counts, and mass estimates will be difficult to
quantify. The most recent studies \citep{coh07} using the Millennium
Simulation and a red-sequence cluster finder are still hampered by the
inability of simulations to reproduce the observed colors of cluster
galaxies. Extensive spectroscopy of at least a subset of clusters
found using optical and X-ray surveys is needed to understand the
contamination rates, redshift distributions, substructure, and other
cluster properties; such work is only now under way \citep{gil07}.

Given that such projections can only enhance the detectability of
lower mass systems, as well as inflate velocity dispersions and
increase the weak lensing signal, mass estimates for single bound
cluster components using such techniques will always tend to be
overestimated. A striking example is Cluster A (Cl1604+4304), whose
initial velocity dispersion estimate was 1200 km s$^{-1}$
\citep{pos01}. Adding many more spectra, but not tightly restricting
the area sampled, \citet{gal04} found $\sigma=700$ km s$^{-1}$. We now
find that the velocity dispersion can be as low as $\sigma=532$ km
s$^{-1}$, more than a factor of two below the initial
measurement. Comparison of velocity dispersions derived using a
variety of galaxy subpopulations (as done in \S4.3) with simulations
is necessary to determine the optimal mass tracer. 
 
\Citet{hoe03} suggested that structures associated with, but not bound
to, a cluster can increase errors in lensing mass estimates
twofold. Our data shows that such filaments may be common at
$z\sim1$. Furthermore, completely independent structures along the
line-of-sight can be cause for concern, even at low redshift
\citep{lok06}. In Cl1604, we find moderate-mass groups at $z\sim1.2$
directly projected onto clusters at $z\sim0.9$. Both types of
structures can affect the derived mass profiles. Typically, NFW
profiles \citep{nfw96} are fit to clusters in both observations and
simulations; clusters have been observed with NFW concentrations {\em
c} incompatible with simulated clusters. Very recent simulations of
weak lensing by clusters plus foreground group-size halos
\citep{kin07} demonstrates that line-of-sight contamination by even
low-mass systems can both decrease {\em} and increase the measured
value of {\em c}, depending on the geometry. It is not yet known how
much this can affect the overall distribution of observed
concentrations and masses from weak lensing. Implementing techniques
that incorporate multi-wavelength data (X-ray, Sunyaev-Z'eldovich,
optical) can overcome some of these issues, but requires both
extensive data and computational power \citep{mah07}, limiting it to
rich, low-redshift clusters or only the most massive structures at
higher redshift for the time being. With somewhat less data, it is
possible to compare X-ray and weak lensing masses for clusters. Recent
work using observational data has shown moderate agreement between
these two techniques, but discrepancies are plentiful and systematic
errors are much too large for precision cosmological measurements
\citep{seh07}. In contrast, some lensing simulations claim that high
completeness, low contamination, and good mass fidelity can be
achieved with weak lensing \citep{pac07}, but this has yet to be
proven with large samples of real data.

Weak lensing has already been used to estimate the mass of Cl1604+4304
\citep[Cluster A,][]{mar05}, resulting in a mass of
$3.67\pm1.47\times(R/500kpc) 10^{14} M_{\odot}$, corresponding to a
velocity dispersion of $1009\pm199$ km s$^{-1}$. This is significantly
higher than the $575^{+110}_{-85}$ km s$^{-1}$ estimated from the
cluster's X-ray temperature of $T_X=2.51^{+1.05}_{-0.69}$ keV
\citep{lub04}. The velocity dispersion presented here lies in between
these two estimates, but is much more consistent with the X-ray
temperature. Earlier work on a much smaller region using HST WFPC2
data suggested multiple mass peaks associated with this structure
\citep{ume00}. Cl1604+4304 is the most isolated component of the
supercluster, with the only foreground structures being a group at
$z=0.495$ and a possible wall at $z=0.600$. Nevertheless, its
estimated mass varies by a factor of nearly two. Improved X-ray
measurements from deep Chandra observations \citep{koc08} and lensing
data from a recent ACS mosaic, combined with our spectroscopic
database, may help resolve this discrepancy.

Instead of measuring individual cluster masses, one could avoid some
of the above problems with a sufficiently large and deep survey by
calculating the mass on very large ($\sim100$ Mpc) scales over several
superclusters, and comparing the results to simulations and structure
formation theories, but such surveys remain in the future. With
excellent deep photometric data, it is also possible to use weak
lensing tomography, incorporating photometric redshifts to map all the
galaxies along the line-of-sight. However, even the best photometric
redshifts have errors which are equal to or greater than the entire
redshift depth of the Cl1604 structure \citep{mar07}. As a result,
uncertainties in the bias and scatter of photo-$z$ estimators with
redshift remain a problem \citep{ma06}, and will require extensive
spectroscopy to calibrate.

Finally, we note that Cl1604 is an optically selected large scale
structure. To whatever extent it is not representative of all
structures at similar redshifts, any conclusions based on Cl1604 alone
may be biased. For this reason, the ORELSE survey has included a
variety of structures originally detected using X-ray, optical and
even radio techniques. Once the survey is complete, we can address
biases due to the cluster detection technique. Within Cl1604, the
supercluster members were targeted based on their red colors (see
\S3), so that the blue galaxy population may be undersampled;
nevertheless, $\sim50\%$ of the targets are outside the red sequence.

\subsection{Future Work}

The extensive spectroscopic database described here is being used for
a variety of projects. Detailed analysis of line strengths as a
function of position in the structure, local density, and host cluster
properties are being derived by stacking spectra \citep{lem08b}. The
redshift range of the supercluster along with our spectral setup
forces the use of the [OII] doublet at 3727\AA~ as a proxy for
H$\alpha$ emission and thus ongoing star-formation activity. Since
[OII] has been shown to be a flawed tracer \citep{yan06}, we are using
near-IR spectroscopy to search for H$\alpha$ emission in a subset of
our confirmed supercluster members in order to determine the
contribution of non-stellar (i.e. nuclear) sources to the observed
[OII] emission. We have also serendipitously detected a significant
number of $z>4$ Ly$\alpha$ sources in the DEIMOS data, and their
properties will be described in \citet{lem08a}.

A 17-pointing ACS mosaic of the field is being used to examine the
CMDs of individual clusters, in the manner of \citet{hom06} and
\citet{and07}, who used two earlier pointings centered on Clusters A
and D. These images also allow us to compare galaxy colors and
morphologies with their spectroscopic properties, and relate them to
their location within the supercluster.  Deep Chandra observations are
being used to map the distribution of AGN within the supercluster and
determine the X-ray luminosity (and hence mass) of individual clusters
in the complex \citep{koc08}.  Our spectroscopic database, along with
our ACS imaging, will allow us to examine the spectral and
morphological properties of AGN host galaxies, as well as their local
environments, in order to shed light on the mechanisms which trigger
nuclear activity in these systems.  Deep VLA 1.4GHz mapping of the
field shows both bright AGN-like sources and an excess of faint,
possibly star-bursting galaxies \citep{mil08}.  Preliminary results
already suggest signs of environmental segregation, with the
radio-loud population found in dense regions which the radio-faint and
X-ray detected AGN avoid.

Deep Spitzer IRAC and MIPS maps have been
obtained, providing both rest-frame K-band luminosities for estimating
the stellar mass content of member galaxies, and search for obscured
star formation that is becoming more prevalent at higher
redshifts. Ground-based K-band imaging is also being used for stellar
mass estimation. Combining all of these observations will allow
detailed SED fitting and modeling of the star formation histories of
supercluster members. For galaxies without spectroscopy, the extensive
multi-wavelength data will be used to estimate redshifts
photometrically, both as input to lensing models and to better map the
supercluster structure.

\section{Conclusions}

We have presented the results of an extensive spectroscopic campaign
to map the Cl 1604 supercluster at $z\sim0.9$. With over 400
confirmed members, this database provides coverage comparable to
studies of local structures, but at a lookback time of 7 Gyr. The
large number of redshifts is crucial for confirming the individual
clusters, measuring accurate velocity dispersions, and finding
contamination due to fore- and background clusters. A
three-dimensional map of the supercluster is constructed, making
possible the comparison of such structures to large simulations. We
also report the discovery of a poor cluster at $z=1.207$ with at least
15 confirmed members within a $\pm600$ km s$^{-1}$ window and $1 h^{-1}_{70}$ Mpc radius, making it
one of the highest redshift structures selected spectroscopically.

This dataset forms the foundation for a variety of follow-up studies,
using many wavelengths to identify preferred sites for inducing and
quenching star formation and AGN activity. The extensive spectroscopy
allows identification of samples of rare objects, including radio and
X-ray sources within the supercluster. Without the large number of
redshifts, it is impossible to definitively associate individual
objects with the supercluster due to the many sources along the line
of sight that are not physically associated with the structure. As an
example, \citet{koc08} and \citet{mil08} show that X-ray and radio
sources within the supercluster are a very small fraction of the total
number of such sources in the field.  The same is true for infrared
sources such as starbursts. Construction of luminosity functions for
these rarer sources has traditionally hampered by the lack of
redshifts \citep{bra06}; our survey is designed to alleviate these
problems. The only comparable project is the PISCES program, with deep
optical imaging and spectroscopy of high-$z$ clusters with Subaru, but
their targets are typically more massive clusters and their
spectroscopic followup is much less complete. Field surveys such as
GOODS, AEGIS and COSMOS provide similar data to ORELSE, both in terms
of multi-wavelength and spectroscopic coverage, but lack the sampling
of dense environments. For instance, COSMOS contains just one large
scale structure at $z\sim0.7$ \citep{guz07}.

Such detailed mapping, encompassing large ($>10$ Mpc) areas will be
essential for a statistical sample of clusters over a wide redshift
baseline to understand the diversity of structures they are embedded
in. The redshift depth of the supercluster is comparable to
photometric redshift errors achievable from optical ground based
surveys, and especially two-band cluster surveys, potentially
enhancing its detectability. At such high redshifts, photometric
errors are larger for faint galaxies, photometric redshift errors
increase, the change in color of the red sequence with unit $\Delta z$
is reduced \citep[see Fig. 1 of][]{gla05}, and clusters are still in
the process of merging and accreting their galaxies. All of these
effects complicate our ability to reliably construct cluster catalogs
over a wide range of redshifts.  Future programs relying on
photometric redshifts, weak lensing, S-Z, and X-ray data (in
particular, luminosities) to estimate cluster masses must understand
the biases inherent in each technique, especially when using large
samples to constrain cosmological parameters. While large cosmological
simulations can provide insight into these issues, their current
inability to accurately reproduce galaxy properties, especially
colors, means that more observations are necessary. The ORELSE survey
will provide such data on a variety of structures at $0.6<z<1.2$,
forming an invaluable resource for both galaxy evolution and
cosmological studies.

\acknowledgements

This material is based upon work supported by the National Aeronautics
and Space Administration under Award No. NNG05GC34ZG for the Long Term
Space Astrophysics Program. Support for program HST-GO-08560.05-A was
provided by NASA through a grant from the Space Telescope Science
Institute, which is operated by the Association of Universities for
Research in Astronomy, Inc., under NASA contract NAS 5-26555.  The
spectrographic data presented herein were obtained at the W.M. Keck
Observatory, which is operated as a scientific partnership among the
California Institute of Technology, the University of California and
the National Aeronautics and Space Administration. The Observatory was
made possible by the generous financial support of the W.M. Keck
Foundation. The authors wish to recognize and acknowledge the very
significant cultural role and reverence that the summit of Mauna Kea
has always had within the indigenous Hawaiian community.  We are most
fortunate to have the opportunity to conduct observations from this
mountain. The observing staff, telescope operators, and instrument
scientists at Keck provided a great deal of assistance. We would also
like to thank the DEEP2 team for making their software available to
us, and especially M. Cooper for his assistance with implementation.

\clearpage


\clearpage

\begin{deluxetable}{llrrcrrlrrrlrrrl}
\tabletypesize{\scriptsize}
\tablecolumns{16}
\tablewidth{0pc}
\tablecaption{Cl1604 Cluster Coordinates, Redshifts and Velocity Dispersions}
\tablehead{
\colhead{}  & \colhead{}  &  \multicolumn{2}{c}{J2000.0} & \colhead{} & \multicolumn{3}{c}{Within $0.5 h^{-1}_{70}$ Mpc} & \colhead{} &  \multicolumn{3}{c}{Within $1.0 h^{-1}_{70}$ Mpc} & \colhead{} &  \multicolumn{3}{c}{Within $1.5 h^{-1}_{70}$ Mpc} \\
\cline{3-4} \cline{6-8} \cline{10-12}  \cline{14-16}\\[-6pt]
\colhead{ID} & \colhead{Name} & \colhead{RA} & \colhead{Dec} & \colhead{}&\colhead{$N$} & \colhead{$z_{mean}$}  & \colhead{$\sigma~(N_{agree})$} & \colhead{} & \colhead{$N$} & \colhead{$z_{mean}$}  & \colhead{$\sigma~(N_{agree})$} & \colhead{} & \colhead{$N$} & \colhead{$z_{mean}$}  & \colhead{$\sigma~(N_{agree})$}
}
\startdata
A & Cl1604+4304 & 241.08946 & 43.07613 & & 21 & 0.89838 & 532$\pm$127 (2)  & & 32 & 0.89861 & 619$\pm$96 (3)  & & 36 & 0.89832 & 682$\pm$109 (3) \\
B & Cl1604+4314 & 241.09890 & 43.23550 & & 20 & 0.86574 & 756$\pm$83 (3)   & & 32 & 0.86531 & 811$\pm$76 (3)  & & 47 & 0.86569 & 767$\pm$76 (3) \\
C & Cl1604+4316 & 241.03162 & 43.26313 & & 8 &  0.93451 & 439$\pm$256 (1)  & & 18 & 0.93511 & 313$\pm$41 (3)  & & \tablenotemark{a} & \nodata & \nodata \\
D & Cl1604+4321 & 241.13865 & 43.35343 & & 28 & 0.92351 & 559$\pm$171 (2)  & & 53 & 0.92280 & 590$\pm$112 (2) & & 69 & 0.92340 & 668$\pm$103 (3) \\
E & Cl1604+4314B & 241.02815 & 43.23343 & & \tablenotemark{b} & \nodata & \nodata & & \nodata & \nodata & \nodata & & \nodata & \nodata & \nodata \\
F & Cl1605+4322 & 241.21314 & 43.37091 & & 7 & 0.93659 & 142$\pm$297 (1)   & &  9 & 0.93608 & 173$\pm$220 (1) & & 13 & 0.93532 & 435$\pm$129 (2) \\
G & Cl1604+4324 & 240.93754 & 43.40520 & & 6 & 0.89932 & \tablenotemark{c} & & 13 & 0.90075 & 409$\pm$101 (3) & & 21 & 0.90210 & 477$\pm$152 (3) \\
H & Cl1604+4322 & 240.89648 & 43.37309 & & 6 & 0.85292 & \tablenotemark{c} & & 10 & 0.85226 & 302$\pm$64 (3)  & & 11 & 0.85243 & 312$\pm$57 (3) \\
I & Cl1603+4323 & 240.79691 & 43.39176 & & 5 & 0.90230 & \tablenotemark{c} & &  7 & 0.90238 & 359$\pm$140 (3) & &  7 & 0.90238 & 359$\pm$140 (3) \\
J & Cl1604+4331 & 240.91905 & 43.51648 & & \tablenotemark{d} &  \nodata & \nodata & & \nodata & \nodata & \nodata & & \nodata & \nodata & \nodata 
\enddata
\tablenotetext{a}{Overlaps cluster B at this radius}  \tablenotetext{b}{No dispersions measured as it overlaps clusters B and C} \tablenotetext{c}{Insufficient redshifts to compute dispersions} \tablenotetext{d}{No spectroscopic coverage}
\label{candinfo}
\end{deluxetable}

\clearpage
\begin{deluxetable}{lrrccrrr}
\tablecolumns{8}
\tablewidth{0pc}
\tablecaption{Fore- and Background Redshift Peaks}
\tablehead{\colhead{ID}  & \colhead{RA (J2000)} & \colhead{Dec (J2000)} & \colhead{Criteria} & \colhead{$z$}  & \colhead{N$_{A}$} & \colhead{N$_{B}$} & \colhead{N$_{C}$} }
\startdata
i & 241.13561    & 43.37704 & C   & 0.459 & \nodata & \nodata & 5 \\
ii & 241.05078   & 43.12914 & AC  & 0.470 & 17 & \nodata & 5  \\
iii & 241.09854   & 43.05963 & ABC & 0.496 & 22 & 5 & 6 \\
iv &241.02019     & 43.24290 & A   & 0.544 & 17 & \nodata & \nodata  \\
v & 240.98975   & 43.34832 & C   & 0.568 & \nodata & \nodata & 5 \\
vi & 241.07777  & 43.26100 & ABC & 0.599 & 39 & 5 & 9 \\
vii & 241.11231 & 43.38718 & ABC & 0.620 & 16 & 5 & 8 \\
viii & 240.94653   & 43.39078 & AC  & 0.642 & 15 & \nodata & 5 \\
ix & 241.14468    & 43.38072 & ABC & 0.696 & 21 & 6 & 11 \\
x &  240.87280  & 43.40267 & AC  & 0.730 & 32 & \nodata & 9 \\
xi & 241.06424  & 43.19683 & AC  & 0.777 & 38 & \nodata & 9 \\
xii & 240.94727 & 43.39270 & ABC & 0.789 & 16 & 5 & 8 \\
xiii & 241.08856  & 43.30186 & A   & 0.809 & 15 & \nodata & \nodata \\
xiv & 241.09376   & 43.33907 & A  & 0.822 & 21 & \nodata & \nodata \\
xv & 241.09265  & 43.06344 & A  & 0.829 & 15 & \nodata & \nodata \\
xvi & 241.06491 & 43.26877 & A   & 0.971 & 18 & \nodata & \nodata \\
xvii & 241.05082& 43.30216 & A   & 0.982 & 24 & \nodata & \nodata \\
xviii & 241.09734  & 43.32171 & AC  & 1.032 & 22 & \nodata & 5 \\
xix & 241.05106 & 43.30613 & ABC & 1.177 & 22 & 5 & 8 \\
xx & 241.13109& 43.35544 & A   & 1.207 & 15 & \nodata & \nodata \\
\enddata
\label{losinfo}
\end{deluxetable}

\clearpage

\clearpage

\clearpage

\clearpage

\clearpage

\clearpage

\clearpage

\clearpage

\clearpage

\clearpage

\clearpage

\clearpage

\clearpage
\clearpage
{\plotone{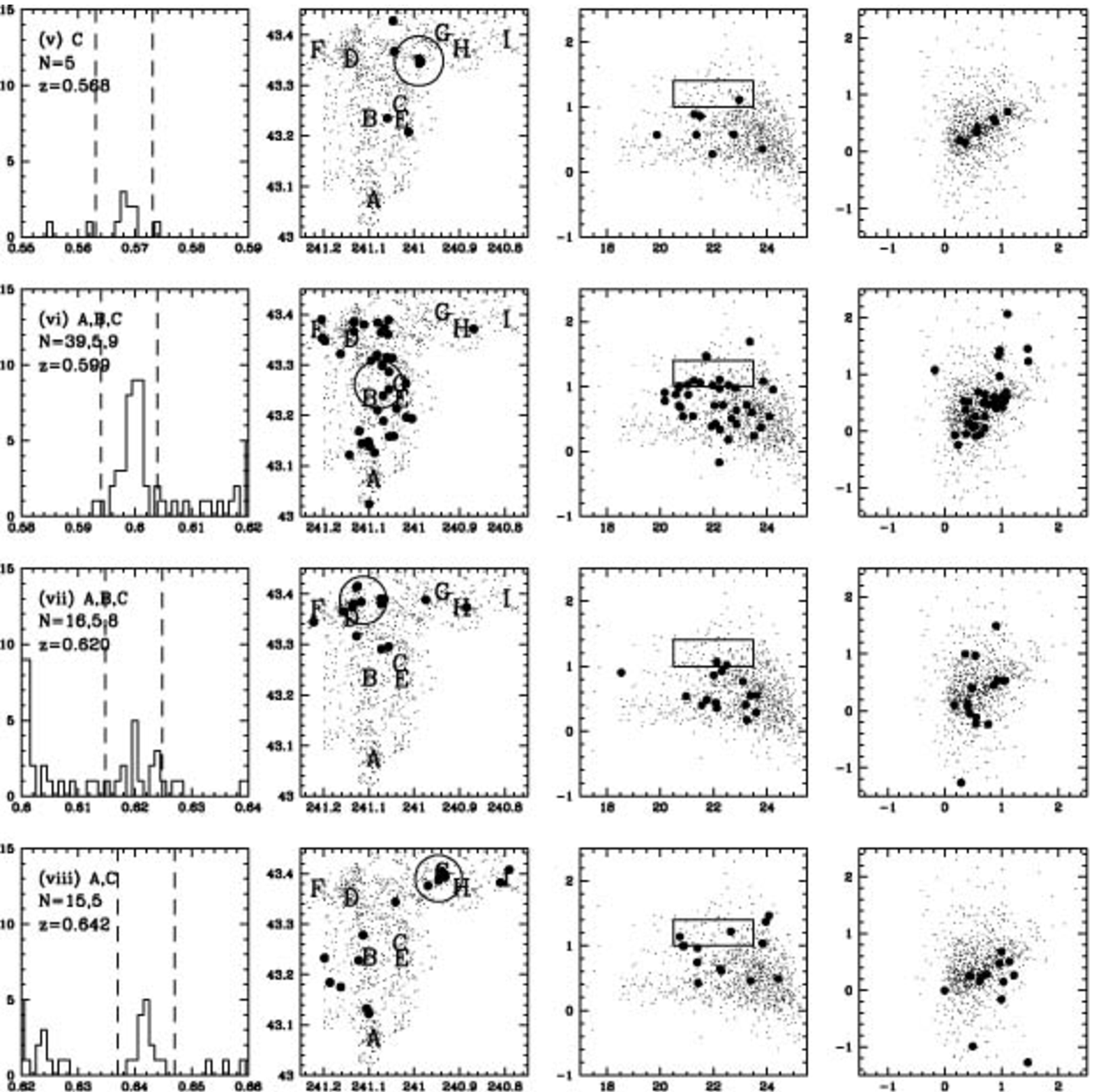}}\\
\centerline{Fig. 13. --- Continued.}
\clearpage
{\plotone{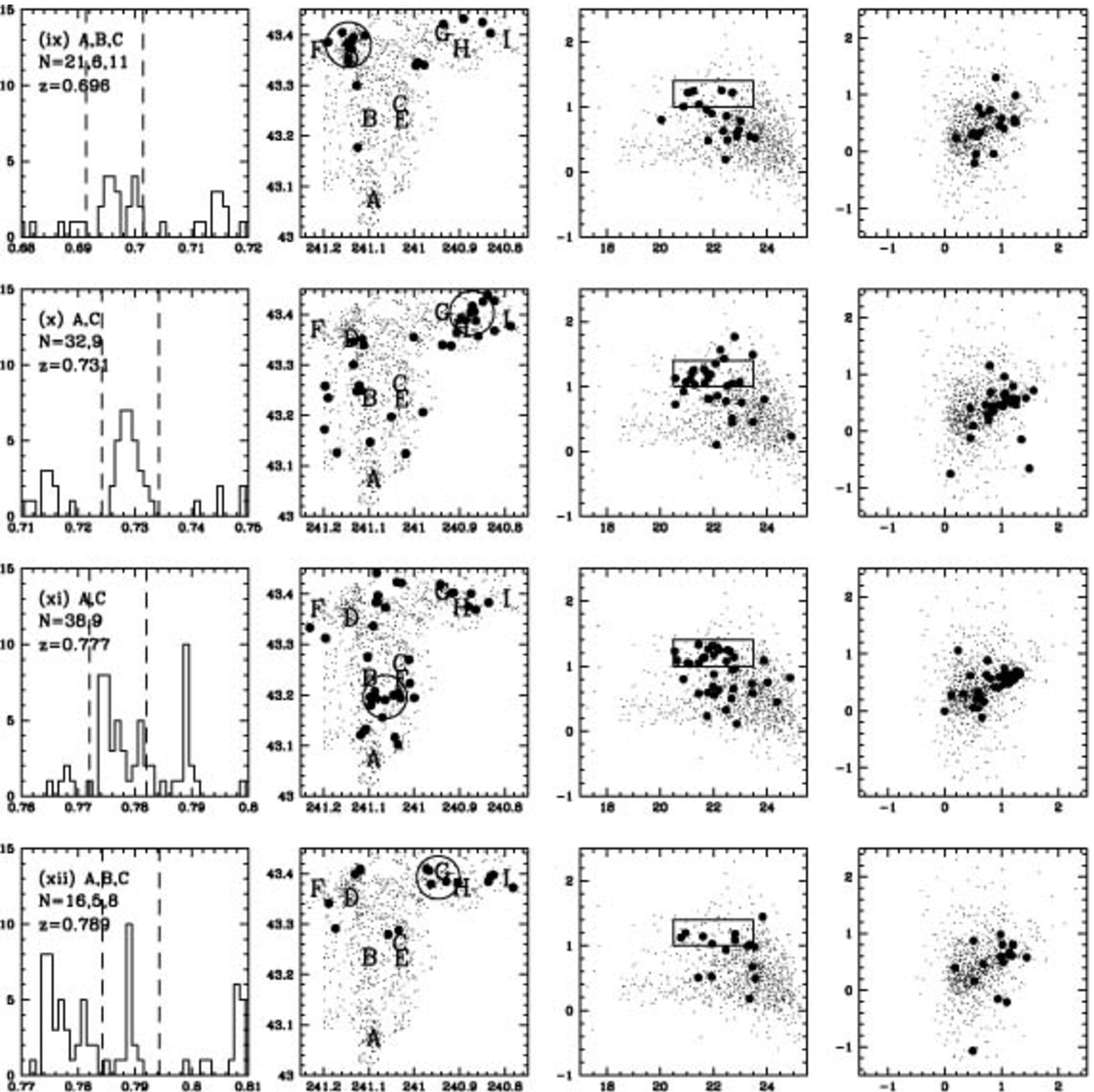}}\\
\centerline{Fig. 13. --- Continued.}
\clearpage
{\plotone{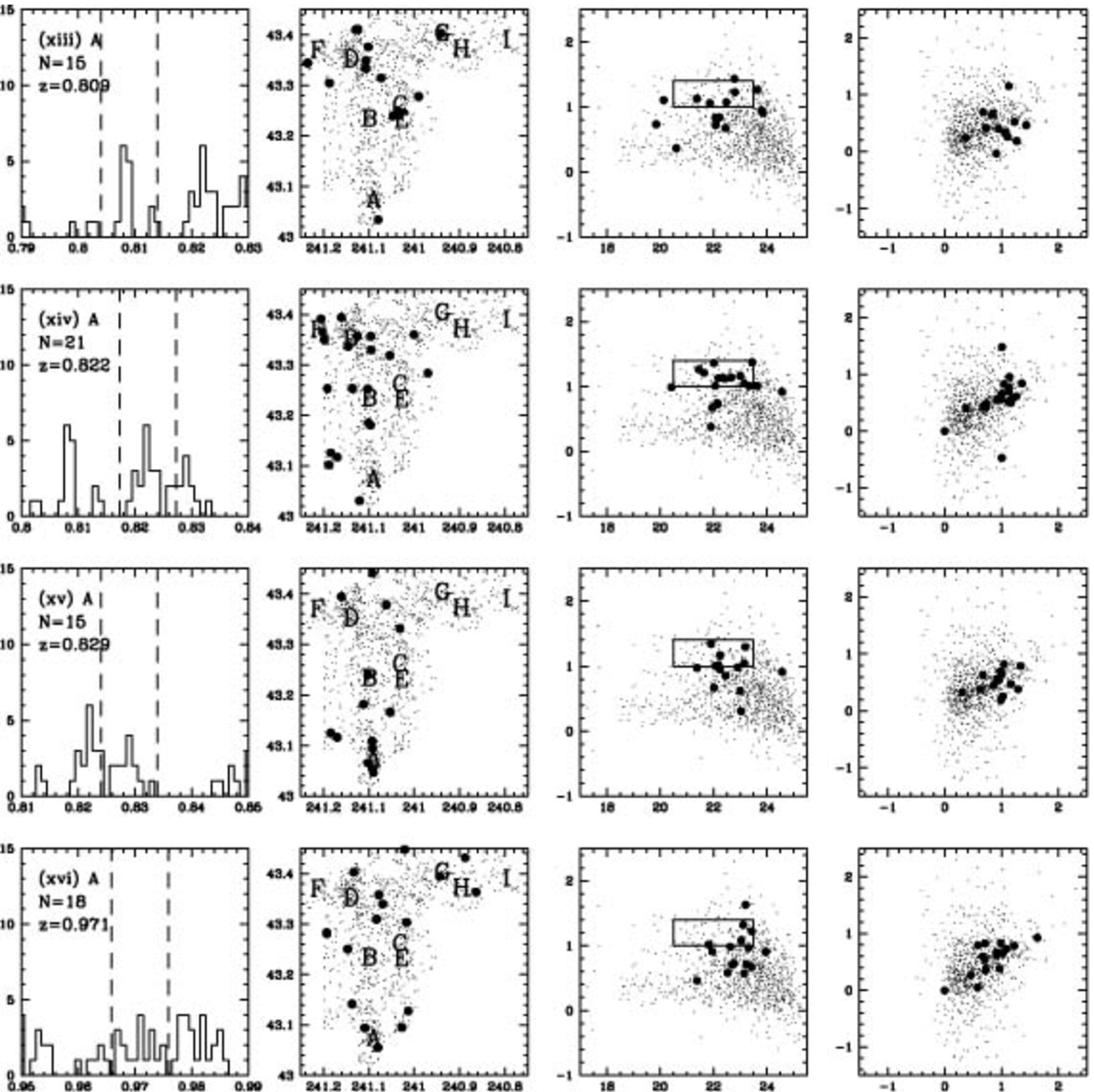}}\\
\centerline{Fig. 13. --- Continued.}
\clearpage
{\plotone{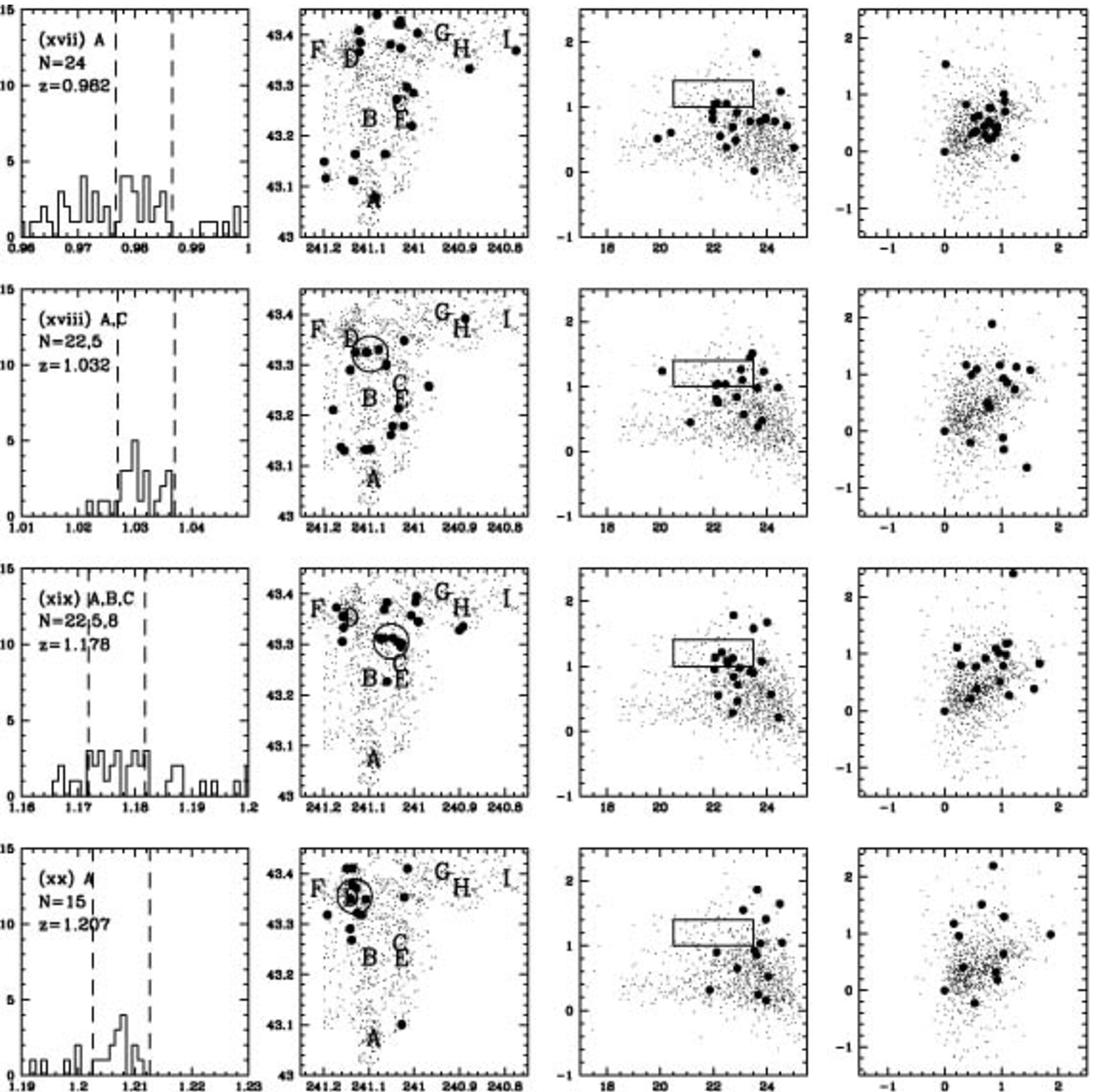}}\\
\centerline{Fig. 13. --- Continued.}


\begin{thebibliography}{foo}

\bibitem[Adelman-McCarthy et al.(2008)]{ade07} Adelman-McCarthy, J.~K., et al.\ 2008, \apjs, 175, 297 

\bibitem[Andreon(2008)]{and07}  Andreon, S.\ 2008, \mnras, 386, 1045 

\bibitem[Beers, Flynn, \& Gebhardt(1990)]{bee90} Beers, T.~C., Flynn, K., \& Gebhardt, K.\ 1990, \aj, 100, 32 

\bibitem[Bertin \& Arnouts(1996)]{ber96} Bertin, E.~\& Arnouts, S.\ 1996, \aaps, 117, 393

\bibitem[Bower et al.(1992)]{bow92} Bower, R.~G., Lucey, J.~R., \& Ellis, R.~S.\ 1992, \mnras, 254, 589 

\bibitem[Branchesi et al.(2006)]{bra06} Branchesi, M., Gioia, 
I.~M., Fanti, C., Fanti, R., \& Perley, R.\ 2006, \aap, 446, 97 

\bibitem[Bruzual \& Charlot(2003)]{bru03} Bruzual, G., \& Charlot, S.\ 2003, \mnras, 344, 1000

\bibitem[Cohn et al.(2007)]{coh07} Cohn, J.~D., Evrard, A.~E., White, M., Croton, D., \& Ellingson, E.\ 2007, \mnras, 382, 1738

\bibitem[Cooper et al.(2007)]{coo07} Cooper, M.~et al.\ 2007, {\em in prep}

\bibitem[Davis et al.(2007)]{dav07} Davis, M.~et al.\ 2007, {\em in prep}

\bibitem[Davis et al.(2003)]{dav03} Davis, M., et al.\ 2003, \procspie, 4834, 161 

\bibitem[De Filippis et al.(2005)]{def05} De Filippis, E., Sereno, M., Bautz, M.~W., \& Longo, G.\ 2005, \apj, 625, 108 

\bibitem[Dressler(1980)]{dre80} Dressler, A.\ 1980, \apj, 236, 351 

\bibitem[Dressler \& Shectman(1988)]{dre88} Dressler, A., \& Shectman, S.~A.\ 1988, \aj, 95, 985 

\bibitem[Dressler et al.(2004)]{dre04} Dressler, A., Oemler, A.~J., Poggianti, B.~M., Smail, I., Trager, S., Shectman, S.~A., Couch, W.~J., \& Ellis, R.~S.\ 2004, \apj, 617, 867 

\bibitem[Evrard et al.(2002)]{evr02} Evrard, A.~E.~et al.\ 2002, \apj, 573, 7

\bibitem[Faber et al.(2003)]{fab03} Faber, S.~M.~et al.\ 2003, \procspie, 4841, 1657

\bibitem[Ford et al.(2003)]{for03} Ford, H.~C., et al.\ 2003, \procspie, 4854, 81 

\bibitem[Gal et al.(2008)]{gal08} Gal, R.~R., Lubin, L.~M., Lemaux,  B.~C., Kocevski, D. \& Squires, G.~K.\ 2008, {\em in prep}

\bibitem[Gal et al.(2005)]{gal05} Gal, R.~R., Lubin, L.~M., \& Squires, G.~K.\ 2005, \aj, 129, 1827 

\bibitem[Gal \& Lubin(2004)]{gal04} Gal, R.~R.~\& Lubin, 
L.~M.\ 2004, \apj, 607, L1 

\bibitem[Gal et al.(2003)]{gal03} Gal, R.~R., de Carvalho, 
R.~R., Lopes, P.~A.~A., Djorgovski, S.~G., Brunner, R.~J., Mahabal, A., \& 
Odewahn, S.~C.\ 2003, \aj, 125, 2064 

\bibitem[Gilbank et al.(2008)]{gil08} Gilbank, D.~G., Yee, 
H.~K.~C., Ellingson, E., Hicks, A.~K., Gladders, M.~D., Barrientos, L.~F., 
\& Keeney, B.\ 2008, \apjl, 677, L89 

\bibitem[Gilbank et al.(2007)]{gil07} Gilbank, D.~G., Yee, H.~K.~C., Ellingson, E., Gladders, M.~D., Barrientos, L.~F., \& Blindert, K.\ 2007, \aj, 134, 282 

\bibitem[Gilbank et al.(2004)]{gil04} Gilbank, D.~G., Bower, 
R.~G., Castander, F.~J., \& Ziegler, B.~L.\ 2004, \mnras, 348, 551

\bibitem[Gladders \& Yee(2000)]{gla00} Gladders, M.~D., \& Yee, H.~K.~C.\ 2000, \aj, 120, 2148 

\bibitem[Gladders \& Yee(2005)]{gla05} Gladders, M.~D., \& Yee, H.~K.~C.\ 2005, \apjs, 157, 1 

\bibitem[Goto et al.(2002)]{got02} Goto, T., et al.\ 2002, 
\aj, 123, 1807

\bibitem[Gunn, Hoessel, \& Oke(1986)]{gho86} Gunn, J.~E., Hoessel, J.~G., \& Oke, J.~B.\ 1986, \apj, 306, 30 

\bibitem[Guzzo et al.(2007)]{guz07} Guzzo, L., et al.\ 2007, 
\apjs, 172, 254 

\bibitem[Hoekstra(2003)]{hoe03} Hoekstra, H.\ 2003, \mnras, 339, 1155 

\bibitem[Holden et al.(1999)]{hol99} Holden, B.~P., Nichol, 
R.~C., Romer, A.~K., Metevier, A., Postman, M., Ulmer, M.~P., \& Lubin, 
L.~M.\ 1999, \aj, 118, 2002 

\bibitem[Homeier et al.(2006)]{hom06} Homeier, N.~L., et al.\ 2006, \apj, 647, 256 

\bibitem[Horne (1986)]{hor86} Horne, K.\ 1986, \pasp, 98, 609

\bibitem[Jannuzi \& Dey(1999)]{jan99} Jannuzi, B.~T., \& Dey, A.\ 1999, Photometric Redshifts and the Detection of High Redshift Galaxies, 191, 111 

\bibitem[Jing \& Suto(2002)]{jin02} Jing, Y.~P., \& Suto, Y.\ 2002, \apj, 574, 538 

\bibitem[Katgert et al.(1996)]{kat96} Katgert, P., et al.\ 1996, \aap, 310, 8 

\bibitem[Kells et al.(1998)]{kel98} Kells, W., Dressler, A., 
Sivaramakrishnan, A., Carr, D., Koch, E., Epps, H., Hilyard, D., \& 
Pardeilhan, G.\ 1998, \pasp, 110, 1487 

\bibitem[King \& Corless(2007)]{kin07} King, L., \& Corless, 
V.\ 2007, \mnras, 374, L37 

\bibitem[Kocevski et al.(2008)]{koc08} Kocevski, D., et al.\ 2008, \apj, {\em submitted}

\bibitem[Kodama et al.(2005)]{kod05} Kodama, T., et al.\ 2005, \pasj, 57, 309

\bibitem[Kodama \& Arimoto(1997)]{kod97} Kodama, T., \& Arimoto, N.\ 1997, \aap, 320, 41 

\bibitem[Lee \& Evrard(2007)]{lee07} Lee, J., \& Evrard, A.~E.\ 2007, \apj, 657, 30 

\bibitem[Lemaux et al.(2008a)]{lem08a} Lemaux, B.~C. et al.\ 2008a, {\em in prep}

\bibitem[Lemaux et al.(2008b)]{lem08b} Lemaux, B.~C., Gal, R.~R, Lubin, L.~M., Kocevski, D. \& Squires, G.~K.\ 2008b, {\em in prep}

\bibitem[{\L}okas et al.(2006)]{lok06} {\L}okas, E.~L., Prada, F., Wojtak, R., Moles, M., \& Gottl{\"o}ber, S.\ 2006, \mnras, 366, L26 

\bibitem[Lubin et al.(2008)]{lub08} Lubin, L.~M., Gal, R.~R, Lemaux, B.~C., Kocevski, D. \& Squires, G.~K.\ 2008, {\em in prep}

\bibitem[Lubin et al.(2004)]{lub04} Lubin, L.~M., Mulchaey, J.~S., \& Postman, M.\ 2004, \apjl, 601, L9 

\bibitem[Lubin et al.(2002)]{lub02} Lubin, L.~M., Oke, J.~B., \& Postman, M.\ 2002, \aj, 124, 1905 

\bibitem[Lubin et al.(2000)]{lub00} Lubin, L.~M., Brunner, R., Metzger, M.~R., Postman, M., \& Oke, J.~B.\ 2000, \apjl, 531, L5 

\bibitem[Lubin et al.(1998)]{lub98} Lubin, L.~M., Postman, M., Oke, J.~B., Ratnatunga, K.~U., Gunn, J.~E., Hoessel, J.~G., \& Schneider, D.~P.\ 1998, \aj, 116, 584 

\bibitem[Ma et al.(2006)]{ma06} Ma, Z., Hu, W., \& Huterer, D.\ 2006, \apj, 636, 21 

\bibitem[Mahdavi et al.(2007)]{mah07} Mahdavi, A., Hoekstra, H., Babul, A., Sievers, J., Myers, S.~T., \& Henry, J.~P.\ 2007, \apj, 664, 62 

\bibitem[Margoniner \& Wittman(2008)]{mar07} Margoniner, V.~E., \& Wittman, D.~M.\ 2008, \apj, 679, 31 

\bibitem[Margoniner et al.(2005)]{mar05} Margoniner, V.~E., Lubin, L.~M., Wittman, D.~M., \& Squires, G.~K.\ 2005, \aj, 129, 20 

\bibitem[Martin \& Sawicki(2004)]{mar04} Martin, C.~L., \& Sawicki, M.\ 2004, \apj, 603, 414 

\bibitem[Martin et al.(2006)]{mar06} Martin, C.~L., Sawicki, M., Dressler, A., \& McCarthy, P.~J.\ 2006, New Astronomy Review, 50, 53

\bibitem[McIntosh et al.(2005)]{mci05} McIntosh, D.~H., Zabludoff, A.~I., Rix, H.-W., \& Caldwell, N.\ 2005, \apj, 619, 193 

\bibitem[Mei et al.(2006)]{mei06} Mei, S., et al.\ 2006, \apj, 639, 81 

\bibitem[Miller et al.(2008)]{mil08} Miller, N., Lubin, L.~M., Gal, R.~R, Lemaux, B.~C., Kocevski, D. \& Squires, G.~K.,\ 2008, {\em in prep}

\bibitem[Navarro et al.(1996)]{nfw96} Navarro, J.~F., Frenk, C.~S., \& White, S.~D.~M.\ 1996, \apj, 462, 563 

\bibitem[Nuijten et al.(2005)]{nui05} Nuijten, M.~J.~H.~M., Simard, L., Gwyn, S., R\"{o}ttgering, H.~J.~A.\ 2005, \apjl, 626, L77

\bibitem[Oke, Postman, \& Lubin(1998)]{oke98} Oke, J.~B., Postman, M., \& Lubin, L.~M.\ 1998, \aj, 116, 549 

\bibitem[Oke et al.(1995)]{oke95} Oke, J.~B., et al.\ 1995, \pasp, 107, 375

\bibitem[Olsen et al.(2005)]{ols05} Olsen, L.~F., Zucca, E., Bardelli, S., Benoist, C., da Costa, L., J{\o}rgensen, H.~E., Biviano, A., \& Ramella, M.\ 2005, \aap, 442, 841 

\bibitem[Pace et al.(2007)]{pac07} Pace, F., Maturi, M., Meneghetti, M., Bartelmann, M., Moscardini, L., \& Dolag, K.\ 2007, \aap, 471, 731 

\bibitem[Paz et al.(2006)]{paz06} Paz, D.~J., Lambas, D.~G., Padilla, N., \& Merch{\'a}n, M.\ 2006, \mnras, 366, 1503 

\bibitem[Pinkney et al.(1996)]{pin96} Pinkney, J., Roettiger, 
K., Burns, J.~O., \& Bird, C.~M.\ 1996, \apjs, 104, 1 

\bibitem[Plionis et al.(2006)]{pli06} Plionis, M., Basilakos, S., \& Ragone-Figueroa, C.\ 2006, \apj, 650, 770 

\bibitem[Postman, Lubin, \& Oke(2001)]{pos01} Postman, M., Lubin, L.~M., \& Oke, J.~B.\ 2001, \aj, 122, 1125

\bibitem[Postman, Lubin, \& Oke(1998)]{pos98} Postman, M., Lubin, L.~M., \& Oke, J.~B.\ 1998, \aj, 116, 560 

\bibitem[Postman et al.(1996)]{pos96} Postman, M., Lubin, L.~M., Gunn, J.~E., Oke, J.~B., Hoessel, J.~G., Schneider, D.~P., \& Christensen, J.~A.\ 1996, \aj, 111, 615

\bibitem[Ramella et al.(2000)]{ram00} Ramella, M., et al.\ 
2000, \aap, 360, 86

\bibitem[Sawicki et al.(2007)]{saw07} Sawicki, M., et al.\ 2007, {\it in prep}

\bibitem[Schlegel et al.(1998)]{sfd98} Schlegel, D.~J., Finkbeiner, D.~P., \& Davis, M.\ 1998, \apj, 500, 525 

\bibitem[Scoville et al.(2007)]{sco07} Scoville, N., et al.\ 2007, \apjs, 172, 150 

\bibitem[Sehgal et al.(2008)]{seh07} Sehgal, N., Hughes, J.~P.,
Wittman, D., Margoniner, V., Tyson, J.~A., Gee, P., \& dell'Antonio,
I.\ 2008, \apj, 673, 163

\bibitem[Sereno et al.(2006)]{ser06} Sereno, M., De Filippis, E., Longo, G., \& Bautz, M.~W.\ 2006, \apj, 645 170 

\bibitem[Silverman(1986)]{sil86} Silverman, B.\ W.\ 1986, 
{\it Monographs on Statistics and Applied Probability}, London: Chapman and Hall 

\bibitem[Simcoe, Metzger, Small, \& Araya(2000)]{sim00} 
Simcoe, R.~A., Metzger, M.~R., Small, T.~A., \& Araya, G.\ 2000, Bulletin 
of the American Astronomical Society, 32, 758

\bibitem[Small et al.(1998)]{sma98} Small, T.~A., Ma, C.-P., Sargent, W.~L.~W., \& Hamilton, D.\ 1998, \apj, 492, 45

\bibitem[Smith et al.(2002)]{smi02} Smith, J.~A., et al.\ 
2002, \aj, 123, 2121

\bibitem[Smith et al.(2005)]{smi05} Smith, G.~P., Treu, T., Ellis, R.~S., Moran, S.~M., \& Dressler, A.\ 2005, \apj, 620, 78 

\bibitem[Swinbank et al.(2007)]{swi07} Swinbank, A.~M., et al.\ 2007, \mnras, 379, 1343 

\bibitem[Tanaka et al.(2004)]{tan04} Tanaka, M., Goto, T., Okamura, S., Shimasaku, K., \& Brinkmann, J.\ 2004, \aj, 128, 2677

\bibitem[Tody(1986)]{tod86} Tody, D.\ 1986, \procspie, 627, 
733 

\bibitem[Tran et al.(2004)]{tra04} Tran, K.-V.~H., Lilly, S.~J., Crampton, D., \& Brodwin, M.\ 2004, \apjl, 612, L89 

\bibitem[Umetsu \& Futamase(2000)]{ume00} Umetsu, K., \& Futamase, T.\ 2000, \apjl, 539, L5 

\bibitem[Visvanathan \& Sandage(1977)]{vis77} Visvanathan, 
N.~\& Sandage, A.\ 1977, \apj, 216, 214

\bibitem[Yan et al.(2006)]{yan06} Yan, R., Newman, J.~A., Faber, S.~M., Konidaris, N., Koo, D., \& Davis, M.\ 2006, \apj, 648, 281 

\bibitem[York et al.(2000)]{yor00} York, D.\ G.\ et al. 2000, \aj, 120, 1579 

\bibitem[Zabludoff \& Franx(1993)]{zab93} Zabludoff, A.~I., 
\& Franx, M.\ 1993, \aj, 106, 1314 

\end{thebibliography}
\end{document}